\documentclass[journal=jacsat,manuscript=article]{achemsoz}
\usepackage[version=3]{mhchem} 
\usepackage{graphicx}
\usepackage{float}
\usepackage{subcaption}
\usepackage{color}
\usepackage{gensymb}
\usepackage{setspace}
\makeatletter
\author{Anne A.Y. Guilbert}
\affiliation{Centre for Plastic Electronics and Department of Physics, Blackett Laboratory, Imperial College London, London SW7 2AZ, United Kingdom}
\email{a.guilbert09@imperial.ac.uk}

\author{Mohamed Zbiri}
\affiliation{Institut Laue-Langevin, 71 avenue des Martyrs, Grenoble Cedex 9, 38042, France}
\email{zbiri@ill.fr}

\author{Peter A. Finn}
\affiliation{Materials Research Institute and School of Biological and Chemical Sciences, Queen Mary University of London, Mile End Road, London E1 4NS, United Kingdom}

\author{Maud Jenart}
\affiliation{Department of Chemistry and Centre for Plastic Electronics, Imperial College London, South Kensington, London SW7 2AZ, United Kingdom}

\author{Peter Fouquet}
\author{Viviana Cristiglio}
\author{Bernhard Frick}
\affiliation{Institut Laue-Langevin, 71 avenue des Martyrs, Grenoble Cedex 9, 38042, France}

\author{Jenny Nelson}
\affiliation{Centre for Plastic Electronics and Department of Physics, Blackett Laboratory, Imperial College London, London SW7 2AZ, United Kingdom}

\author{Christian B. Nielsen}
\affiliation{Materials Research Institute and School of Biological and Chemical Sciences, Queen Mary University of London, Mile End Road, London E1 4NS, United Kingdom}

\makeatletter
\let\acs@address@list\relax
\makeatother

\title{Mapping Microstructural Dynamics up to the Nanosecond of the Conjugated Polymer P3HT in the Solid State}

\begin{document}

\begin{abstract}
We present a detailed microscopic study of the structure-dynamics relationship of both regio-regular (RR) and regio-random (RRa) poly(3-hexylthiophene) (P3HT) using synergistically different elastic, quasi-elastic and inelastic neutron scattering techniques. Deuteration is employed to modulate the coherent and incoherent cross-sections of the materials, beyond a contrast variation purpose, allowing particularly to access both self-motions and collective dynamics of the materials. The neutron scattering measurements are underpinned by extensive quantitative numerical simulations using large-scale classical molecular dynamics (MD) simulations, as well as molecular and periodic first principles quantum chemical (QC) calculations. MD simulations reproduced well the main structural features and slow motions, and shed light on differences in collective dynamics between Q-values linked with the $\pi-\pi$ stacking and the lamellar stacking of the polymer, with the crystalline phase being the most impacted. On the other hand MD led to a limited description of molecular vibrations. In this context, first principles molecular QC calculations described well the high-energy vibrational features ( $>$ 900 cm$^{-1}$ ), while periodic QC allowed to describe the low- and mid-energy vibrational range ( 200-900 cm$^{-1}$ ). The mid-energy range is predominantly associated with both intra-molecular and inter-molecular mode coupling, which encloses information about both the polymer conformation and the polymer packing at short range. We show that the presented combined approach of neutron-based measurements and multi-computational calculations allows to fully map out the structural dynamics of conjugated polymers such as P3HT. One of the outcomes of this study is the validation of the common assumption made that RRa-P3HT is a good approximation for the amorphous phase of RR-P3HT at the macroscopic level, although some differences are shown at the molecular level. The present work helps to clarify unambiguously the latter point which has been largely overlooked in the literature. By comparing the neutron vibrational results with available Raman and IR data in the literature, we highlight the importance to complement such optical spectroscopy techniques with inelastic neutron scattering. The latter offering the advantage of being insensitive to the delocalized $\pi$-electron system, and thus enabling to infer relevant quantities like conjugation lengths, for instance, impacting properties of conjugated polymer.
\end{abstract}

\section{Introduction}
Conjugated polymers have attracted keen interest over the past decade for their potential applications as semiconductors in various types of devices: organic light emitting diodes, organic solar-cells, organic field-effect transistors, etc. Because polymers are soft materials, a range of dynamics occurs over an extended time scale, from femtosecond to millisecond, and are likely to impact the optoelectronic properties of the material.\\
Femtosecond dynamical processes like vibrations have been evidenced to impact absorption,\cite{Clark2007} inner reorganization energy,\cite{Shuai2014} charge transfer between  molecules of the same types\cite{Barbara1996} and between different molecules at an heterojunction,\cite{Falke2014} delocalization,\cite{Nelson2017} and so more generally charge transport\cite{Oberhofer2017} and charge separation processes.\cite{DeSio2017} Slower dynamics, on the picosecond to nanosecond time-scale, include side-chain reorientation and backbone torsion\cite{Guilbert2015}. These dynamics are correlated with the conformations of the chains and thus, impact the efficiency of the optoelectronic processes. It has been shown for instance that the global charge transport network changes over a timescale competing with the charge carrier lifetime.\cite{Jackson2016} Furthermore, these slow dynamics are temperature-dependent and therefore, can be activated during device operation. For instance, molecular diffusion becomes predominant above the glass transition of the material and is a known degradation mechanism.\cite{Bertho2008}\\
While lot of efforts have been invested in studying the microstructure of organic semiconductors at different length-scale, with a focus on understanding the relationship between the microstructure and optoelectronic properties, little has however been done to fully characterize the structural dynamics of organic semiconductors, in order to understand its impact on this relationship. By structural dynamics, we refer to vibrational as well as local (diffusive and rotational) dynamics in such materials. In this paper, we attempt to fully map the structural dynamics of semi-crystalline conjugated polymers. To increase solubility, alkyl side chains are added to the conjugated backbones of the polymers. The side chains are not involved in the frontier molecular orbitals and thus, do not impact directly optoelectronic properties. However, they do impact the polymer conformation and the polymer packing, which in turn impact the optoelectronic processes and the subsequent emergent macroscopic properties of the material. We aim, more specifically, at understanding and decoupling the effects of the backbone, the side chains as well as their conformational distribution in the solid state, on the structural dynamics of the crystalline and amorphous phases of semi-crystalline polymeric semiconductors. We consider the model systems regio-regular P3HT (RR-P3HT).\\
Although RR-P3HT is one of the most studied conjugated polymers, there is no general consensus on the degree of crystallinity of RR-P3HT. This is mainly due to the disorder in the crystalline phase, the possible order in the amorphous phase, and the impact of regioregularity, molecular weight and processing. Wide-angle X-Ray diffraction has often been used to characterize the crystallinity of RR-P3HT. Disorder in polymer crystals caused by thermal vibrations and lattice imperfections lead in general to an underestimation of the degree of crystallinity. Further challenges arise in assessing accurately and correctly the contributions to diffraction of the incoherent scattering of the crystalline and the amorphous components. Regio-random P3HT (RRa-P3HT) is often used as a model of the amorphous phase of RR-P3HT as it has been shown that macroscopic properties such as absorption, photoluminescence\cite{Brown2003} and charge transport\cite{Sirringhaus1999} are varying with the percentage of regioregularity of P3HT. Therefore, we also interrogate the validity of this approximation, which to the best of our knowledge is not discussed in the literature.\\
\begin{table}[h]
\begin{center}
\caption{Bound scattering length (fm) of hydrogen ($^{1}$H) and deuterium ($^{2}$H).}\label{tab:SLD-1}
\begin{tabular}{ccc}
\hline
 & coherent & incoherent \\ \hline
$^{1}$H & -3.7406 & 25.274 \\
$^{2}$H & 6.671 & 4.04\\ \hline
\end{tabular}
\end{center}
\end{table}

Neutron scattering offers the possibility of modulating the coherent and incoherent cross-sections of materials by deuteration (Tables \ref{tab:SLD-1}). In a previous work, done within the framework of organic photovoltaics, we have evidenced a host-guest property exchange in the well studied system poly(3-hexylthiophene) (P3HT):phenyl-C61-butyric acid methyl ester (PCBM) by combining quasi-elastic neutron scattering (QENS) \cite{Guilbert2016} and classical molecular dynamics simulation (MD). \cite{Guilbert2017} The QENS measurements highlighted that P3HT is vitrified upon blending with PCBM, while PCBM is plasticized upon the same blending process with P3HT.\cite{Guilbert2016} MD allowed us to study the different phases of the blend microstructure separately. We showed that this host-guest property exchange occurs within the amorphous phase of the blend and is favoured by partial wrapping of P3HT around PCBM with the thiophene rings co-facial to PCBM cage. The side chains are pushed away to accommodate PCBM.\cite{Guilbert2017} We suggested that this was beneficial for charge transfer from P3HT to PCBM. This has been since studied in details using DFT by Zheng \textit{et al.}\cite{Zheng2019} This highlights the need to fully characterize, down to the molecular level, the amorphous phase of such conjugated polymers.\\
While small-angle neutron scattering technique (SANS) has largely been used to characterize the conformation of conjugated polymers in solution or gels,\cite{Huang2010,McCulloch2013,Li2009,Sobkowicz2012,Newbloom2011} only a few works report on SANS measurements in the solid-state.\cite{Yin2011,Olds2012} Neutron diffraction requires deuterated materials that can be challenging to synthesize, in addition to their prohibitive cost. QENS has been used to primarily characterize the self-motion dynamics of conjugated polymers side chains as it takes advantages of the large number of hydrogens and their pronounced neutron scattering power.\cite{Etampawala2015,Paterno2016,Paterno2013,Zhan2018,Guilbert2015,Wolf2019} However, QENS is still underused as it also requires a sufficient amount of materials. Although inelastic neutron scattering (INS) is also dominated by the hydrogen contribution, it has only been recently used for conjugated polymers \cite{VanEijck2002,Harrelson2017} as the signal is largely reduced in comparison with the elastic component and its neighboring QENS signal.\cite{Cavaye2019}\\
To fully characterize the structural dynamics of our systems, we combined synergistically various neutron scattering techniques, covering a Q-range of 0.04 \r{A}$^{-1}$ to 2.4 \r{A}$^{-1}$ (equivalent to distances of a couple \r{A} to 10s of \r{A}), and a femtosecond to nanosecond time-scale, to probe both fast local vibrational dynamics and slow rotational, transitional and diffusive motions. We used neutron diffraction to probe the wide angle structure, as well as extending the measurements down to the small angle limit. Local and vibrational dynamics are probed using synergistically different neutron spectroscopic techniques. Cold-neutron time-of-flight and spin-echo measurements are performed to probe local and relaxation dynamics by covering both the picosecond and nanosecond time scales. Hot-neutron vibrational spectroscopy is carried out to map out the full vibrational spectrum, including both the external/lattice modes and the internal/molecular degrees-of-freedom (Scheme \ref{sch:intro}). Note that most charge transfer processes in conjugated polymers occur on the femtosecond to picosecond timescale. On the other hand, charge transport occurs through a hopping process, and is characterized by a longer timescale of hundreds of picoseconds to tens of nanoseconds, and resulting from charge transfer involving local energy sites. Thus, we cover largely the relevant timescale to optoelectronic processes. We also make use of one of the most powerful advantages of neutron scattering, consisting of adopting the deuteration technique; deuteration can be used for contrast variation on one hand to "hide" or "enhance" the signal of part of a molecule in the context of this work, but also to study independently coherent and incoherent processes. We particularly present QENS and INS measurements on deuterated samples in order to link microstructure and collective motions or lattice vibrations. Presently, measurements are underpinned by extensive ab-initio and classical MD simulations. We utilized molecular and periodic quantum chemistry methods to model individual defects and monitor their influence on the behavior of the molecule in the solid state. MD is used to monitor the temporal and thermodynamical evolution of the polymer conformations, in the different phases; amorphous or crystalline. By comparing various calculated quantities with neutron scattering experiments, we were able to assess (i) the goodness of RRa-P3HT as a model of the amorphous phase of RR-P3HT, (ii) the efficiency of the computational molecular approach with respect to the periodic solid-state approach, and (iii) various models of RR-P3HT. We place a specific emphasis on the impact of disorder, degrees of freedom and packing, and on investigating the adequacy of the theory needed to describe and to capture the essential structural and dynamical features of these conjugated polymers.\\
\begin{scheme}[H]
\includegraphics[width=0.95\textwidth]{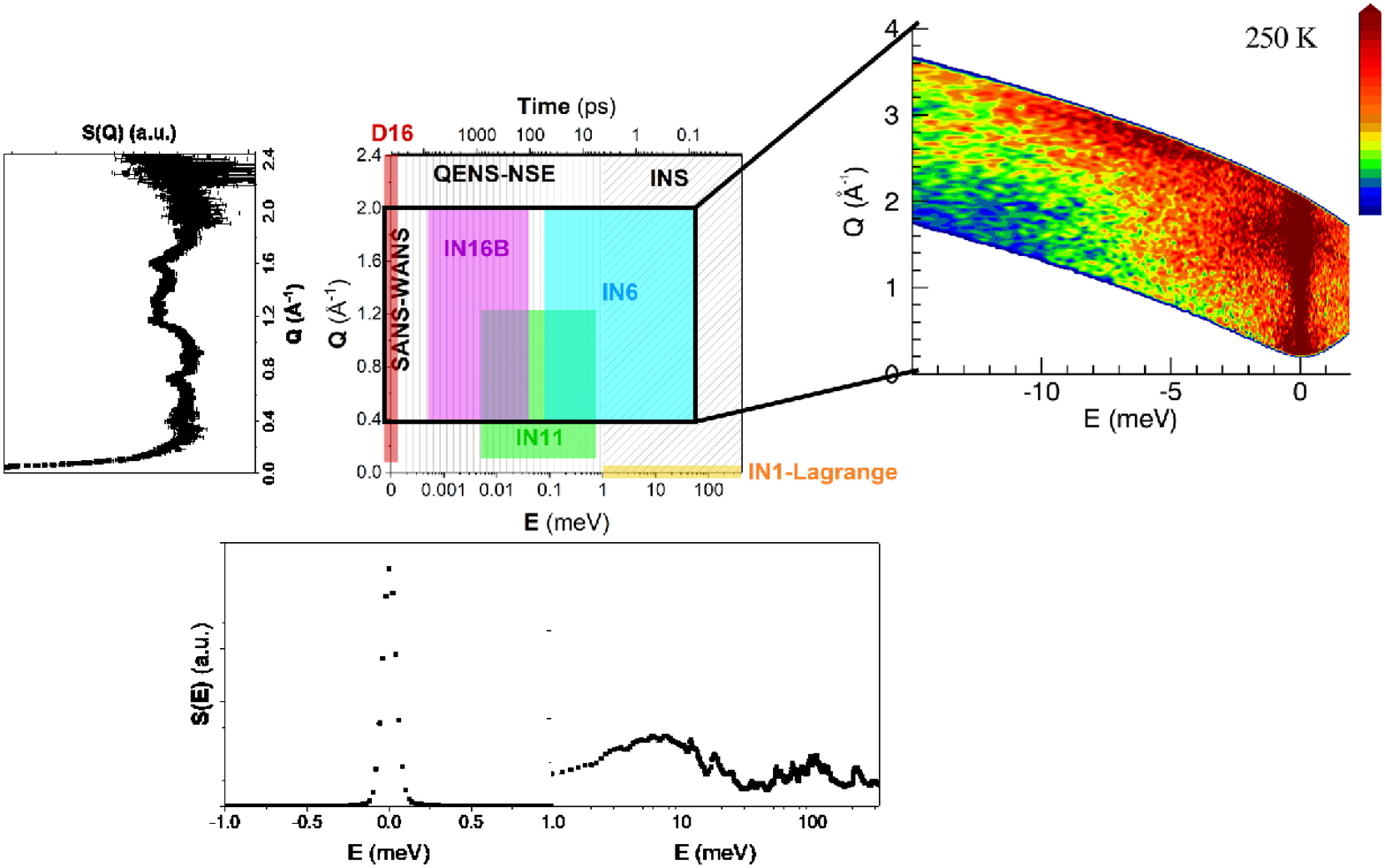}
\caption{Map representing the Q-E space of the instruments used in this work with the S(Q,E) map, shown as an inset, of deuterated RR-P3HT taken at 250K from IN6. Are depicted on the the S(Q,E) the elastic scattering at E=0 meV, a quasielastic (QENS) component between $\sim$ -1 and 1 meV, and the INS for energy below -1 meV. Negative and positive signs of the energy axis indicate the anti-Stokes (neutron energy gain) and Stokes (neutron energy loss) sides. The elastic signal, S(Q), of deuterated RR-P3HT (d-RR-P3HT) taken from D16 at 296K is shown on the left-hand side. The Q-averaged QENS signal, S(E), of d-RR-P3HT from IN6 is depicted on the bottom left-hand side, and the inelastic (INS) signal, S(E), of d-RR-P3HT taken on IN1-Lagrange at 10K, is depicted on the bottom right-hand side.}\label{sch:intro}
\end{scheme}
\section{Results and discussion}
In the following, hydrogenated and deuterated RR/RRa-P3HT are labelled by a prefix h- and d-, respectively. We study both hydrogenated and deuterated RR-P3HT and RRa-P3HT (Table \ref{tab:SLD-2}).\\
\begin{table}[h]
\begin{center}
\caption{Molecular weight (Mw) and polydispersity (PDI) as measured by gel permeation chromatography, regioregularity (RR) as measured by NMR (Figure S7 in Supporting Information) and scattering length density (SLD) calculated assuming a density of 1 g/cm$^{3}$.}\label{tab:SLD-2}
\begin{tabular}{ccccccc}
\hline
 & Mw (in kDa) & PDI & RR (\%) & SLD (in 10$^{-6}$ \r{A}$^{-2}$)\\ \hline
h-RR-P3HT & 34.1 & 1.74 & 94.7 & 0.61417 \\
d-RR-P3HT & 30 & 1.5 & 95 & 5.4344 \\
h-RRa-P3HT & 304 & 3.2 & 56 & 0.61417\\
d-RRa-P3HT &  53 & 1.95 & 73 & 5.4053\\ \hline
\end{tabular}
\end{center}
\end{table}
\subsection{Investigating the Structure}
\begin{figure}[H]
\includegraphics[width=0.8\textwidth]{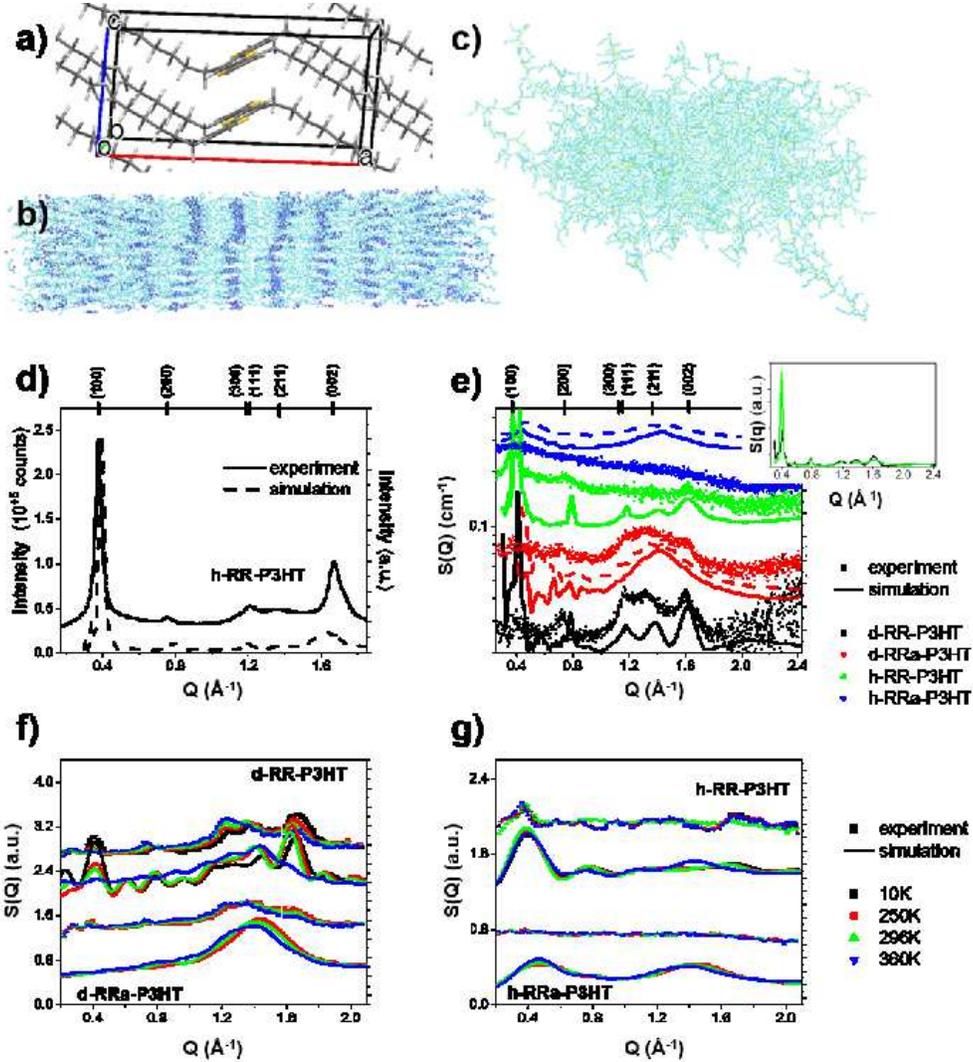}
\caption{(a) Unit cell of the proposed refined crystal structure for RR-P3HT \cite{Kayunkid2010}. (b) MD simulation box of crystalline RR-P3HT after a simulation time of 20 ns at 300K. (c) MD simulation box of amorphous RR-P3HT after a simulation time of 20 ns at 300K. (d) Integrated diffraction intensity from WAXS taken in transmission (solid line) of h-RR-P3HT with peaks assignment following the crystal structure depicted in (a) and calculated XRay spectrum from MD simulation at 300K (dashed line). (e) WANS spectra of d-RR-P3HT, d-RRa-P3HT, h-RR-P3HT and h-RRa-P3HT at 300K from D16 (squares) with peaks assignment following the crystal structure depicted in (a) and calculated neutron spectra from MD simulation at 300K (solid lines). The dashed lines are the calculated neutron spectra from MD simulation on the amorphous mode of RR-P3HT and are thus, compared with RRa-P3HT measurements and calculations. The inset compares the difference in intensity of the (100) reflection calculated from MD simulation for crystalline d-RR-P3HT (black) and h-RR-P3HT (green). Temperature evolution of the neutron diffraction patterns of (f) d-RR-P3HT and d-RRa-P3HT, and (g) h-RR-P3HT and h-RRa-P3HT, extracted from IN6 measurements (squares), compared to calculated patterns from MD simulations (solid lines). The experimental and calculated diffractograms of d-RR-P3HT and h-RR-P3HT have been shifted vertically, with respect to d-RRa-P3HT and h-RRa-P3HT, for clarity.}\label{fig:D16}
\end{figure}

As shown in Figure S8 (a), and in agreement with reference \cite{Shao2014}, the diffraction pattern as measured by X-Ray scattering as well as the absorption spectrum is almost not affected by deuteration. We assign the diffraction peaks observed for d-RR-P3HT and h-RR-P3HT following the refined crystal structure by Kayunkid \textit{et al.}.\cite{Kayunkid2010} The unit cell is displayed in Figure \ref{fig:D16} (a). The usual (100), (200) and (300) reflections located at $\sim$ 0.37 \r{A}$^{-1}$, $\sim$ 0.75 \r{A}$^{-1}$, $\sim$ 1.21 \r{A}$^{-1}$, respectively, correspond to the lamellar stacking. The (020) reflection at $\sim$ 1.67 \r{A}$^{-1}$ is related to the $\pi-\pi$ stacking. All these reflections are observed for both h- and d-RR-P3HT.  On the other hand, in this Q-range, d- and h-RRa-P3HT exhibit two diffuse amorphous halos at $\sim$ 0.4 \r{A}$^{-1}$ and 1.5 \r{A}$^{-1}$, corresponding to the average separation between adjacent alkyl chains and between thiophene backbones, respectively.\cite{Shen2016} It is worth to notice that d-RRa-P3HT exhibits some structural reminiscence in terms of some diffraction peaks observed for h-/d-RR-P3HT, which is not surprising given that d-RRa-P3HT is not fully random (RR=73\%) (Table \ref{tab:SLD-2}).\\
We performed wide-angle neutron scattering (WANS) using the high-resolution diffractometer D16 at the ILL on d-RR-P3HT, d-RRa-P3HT, h-RR-P3HT and h-RRa-P3HT (Figure \ref{fig:D16} (e)). The flat background results from the incoherent scattering of the molecules; thus, the hydrogenated samples have a higher background than the deuterated samples. The (100) reflection is strongly suppressed for d-RR-P3HT in comparison with the X-Ray data. In contrast, the (100) reflection for h-RR-P3HT is very strong while (111) and (211) reflections are completely suppressed. The modulation of the intensities is due to the scattering length density (SLD) of the side chains increasing from -0.39867.10$^{-6}$ \r{A}$^{-2}$ to 5.0073.10$^{-6}$ \r{A}$^{-2}$ upon deuteration, while the SLD of the backbones is varying from 1.0265.10$^{-6}$ \r{A}$^{-2}$ to 1.4248.10$^{-6}$ \r{A}$^{-2}$ (Table \ref{tab:SLD-2}). The partial contribution to the neutron structure factor related to the cross-correlation between backbones and side chains is therefore negative for the hydrogenated samples. Thus, it can be concluded that the (111) and (211) reflections are due to the cross-correlation between backbones and side chains, while (100), (200), (300) and (020) are due to backbone-backbone or side chain-side chain correlations. A broad peak around 1.5 \r{A}$^{-1}$ is observed in d-RRa-P3HT, although some reminiscence of the diffraction peaks of d-RR-P3HT, as observed by X-Ray diffraction, are present due to a reduced regioregularity character. Only two broad peaks around 0.4 and 1.5 \r{A}$^{-1}$ are exhibited by h-RRa-P3HT, which is in agreement with the X-Ray diffraction data (Figure \ref{fig:D16} (d)).

We calculated the WAXS and WANS spectra of different crystalline and amorphous models from MD simulations using the Debye scattering formula (see Figure \ref{fig:D16} (b-c)):\cite{Debye1915}
\begin{equation}
    I(Q) = \sum_{i}\sum_{j}f_{i}f_{j}\frac{sin(Qr_{ij})}{Qr_{ij}}
\end{equation}
where $r_{ij}$ is the distance between two atoms and $f_{i}$ is the atomic form factor of atom $i$. Note that for X-Ray, the atomic form factor is Q-dependent while for neutron, the atomic form factor can be assumed to be Q-independent.\\
The calculated WAXS (Figure \ref{fig:D16} (d)) and WANS (Figure \ref{fig:D16} (e)) agree reasonably well with the experimental data. The modulation of the (100) peak intensity by deuteration is reproduced by the simulation, although unlike in the experiment, the peak is not fully suppressed upon deuteration. The simulation captures the suppression of the (111) and (211) peaks in the hydrogenated samples. Note that the Debye scattering formula assumes an isotropic character of the material, which is not the case for the crystalline form of P3HT. Also, although the MD simulations are of a large scale nature, the modelling boxes are of a finite size. This leads to an increase of the background at low momentum transfers if the periodic boundary conditions are not taking into consideration. In this context the background has been subtracted in the presented calculations (Figure \ref{fig:D16} (d-e)), or interference peaks are present when periodic boundary conditions are applied for the calculation of the patterns (Figure \ref{fig:D16} (f-g)). Interestingly, no differences are observed between RR-P3HT and RRa-P3HT models created from the melt. The end-to-end distance and radius of gyration for RR-P3HT and RRa-P3HT in the samples created from the melt are of the same order of magnitude with a ratio of the mean-square end-to-end distance and mean-square radius of gyration approaching 6; the ratio approaches 10 for the crystalline RR-P3HT sample (Table S1 in the SI). Thus, in the amorphous sample, both RR-P3HT and RRa-P3HT chains can be seen as coils, while in the crystalline phase, RR-P3HT chains can be seen as rigid rods. However, the dihedral distribution for RR-P3HT and RRa-P3HT in the samples created from the melt are different (Figure S16 in the SI) and the simulation box is anisotropic for RR-P3HT, pointing towards a preferential alignment.\\
Diffraction measurements using D16 capture the high-Q part of the small-angle neutron scattering profiles (Figure S9 in the SI). The incoherent background exceeds the coherent signal from the hydrogenated samples, while Porod's scattering is observed for both the  RR and RRa deuterated samples, with a derived fractal exponent around 2.4. A fractal exponent of 2 is indicative of Gaussian chains, while a fractal exponent of 3 is indicative of clustered networks within the framework of the mass fractal model. Note that a fractal exponent between 3 and 4 could point toward an interface (rough if 3 and smooth if 4), within the framework of the surface fractal model. We therefore suggest that, given the complex microstructure of this semi-crystalline polymer, there is likely a presence of Gaussian chains, with a contribution from the interface between the crystalline phase and the amorphous matrix, which in turn would be a rough interface with a fractal exponent of 3.\\
We also extracted the temperature-dependent diffractograms of all the polymers, from the elastic component of the  measurements carried out using the neutron time-of-flight spectrometer IN6 at the ILL (Figure \ref{fig:D16} f-g). For d-RR-P3HT, upon heating, the intensity of the (100) reflection decreases, while the intensity of the (200) and (300) reflection increases. The (002) reflection decreases in intensity (Figure \ref{fig:D16} (f)). As expected, the peaks are shifted to lower Q values with increased temperature due to thermal expansion. For d-RRa-P3HT, a similar trend is observed (Figure \ref{fig:D16} (f)). For h-RR-P3HT, (100) reflection slightly increases in intensity and shifts towards lower Q values above room temperature, consistent with a thermal expansion of the lamellar stacking (Figure \ref{fig:D16} (g)). On the other hand, h-RRa-P3HT remains featureless as temperature changes (Figure \ref{fig:D16} (g)).\\ 
The calculated temperature-dependent WANS spectra agree well with the measurements. There seems to be a transition between 300K and 360K for RR-P3HT, where the peaks reflecting a cross-correlation between backbones and side chains are becoming more pronounced but also less defined. This may be due to a temperature-induced amorphisation of the side chains. In the simulation, the ratio between the mean-square of the end-to-end distance and the mean-square of the radius of gyration increases between 300K and 360K, without reaching the high values observed in the melt.\\
Having probed the temperature dependence of the microstructural aspects of the RR and RRa P3HT, and characterized their signature using elastic scattering techniques (neutrons and x-rays), we move a step further and extend the static picture by exploring dynamics. First, we investigated the time-dependent microscopic behavior of the systems by performing various spectroscopic measurements based on different time/energy-resolved neutron techniques, and we used deuteration to link collective motions to structural features. Second, we used vibrational spectroscopy, in combination with deuteration, allowing us to provide further information on the long-range order, coherent lattice modes, as well as incoherent molecular scattering and short-range order; like segmental orientations.

\subsection{Investigating the Dynamics}
\paragraph*{Quasi-Elastic Neutron Scattering and Neutron Spin-Echo}

\begin{figure}[H]
\begin{subfigure}[b]{0.45\textwidth}
\includegraphics[width=\textwidth]{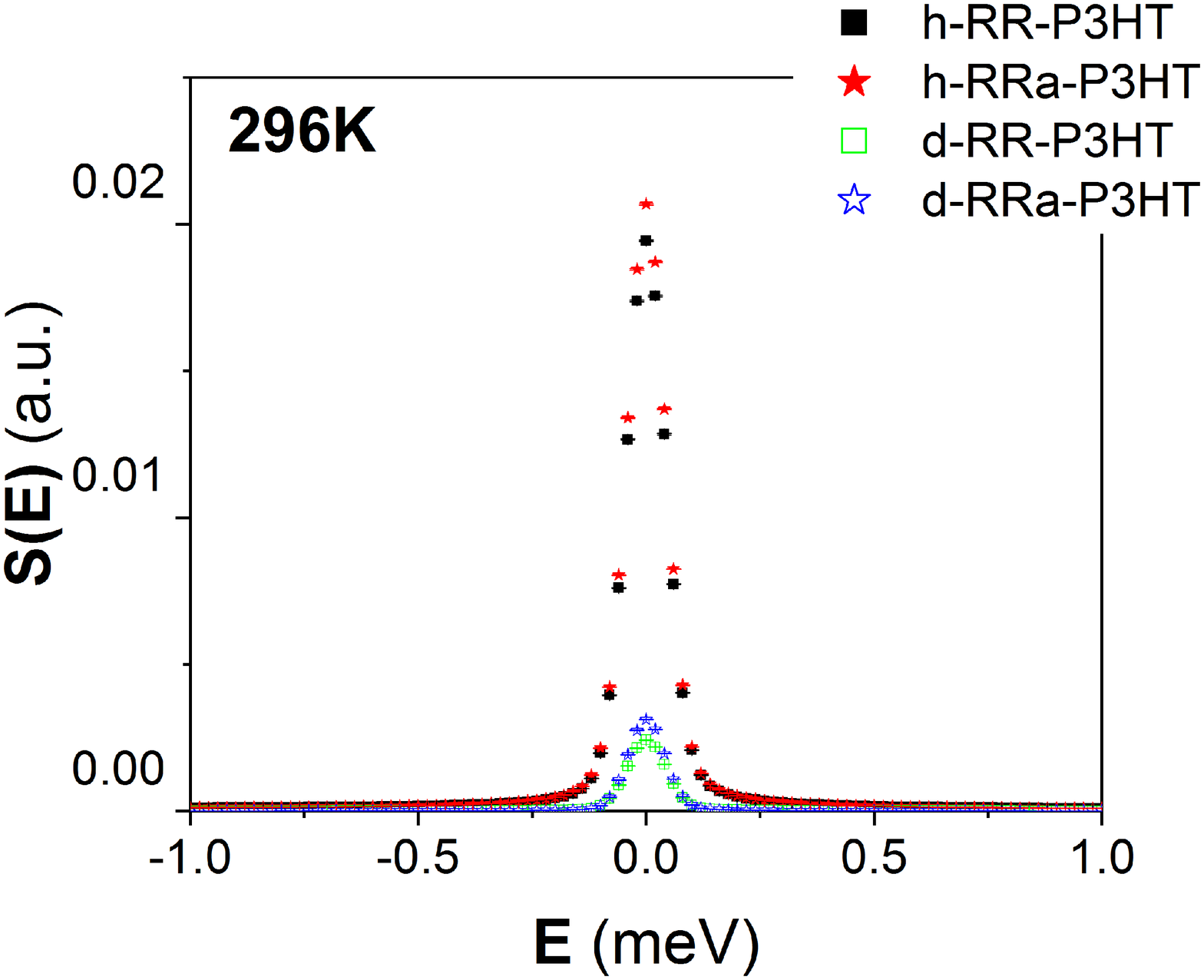}
\caption{}\label{fig:QENS_coh_inc}
\end{subfigure}
\begin{subfigure}[b]{0.45\textwidth}
\includegraphics[width=\textwidth]{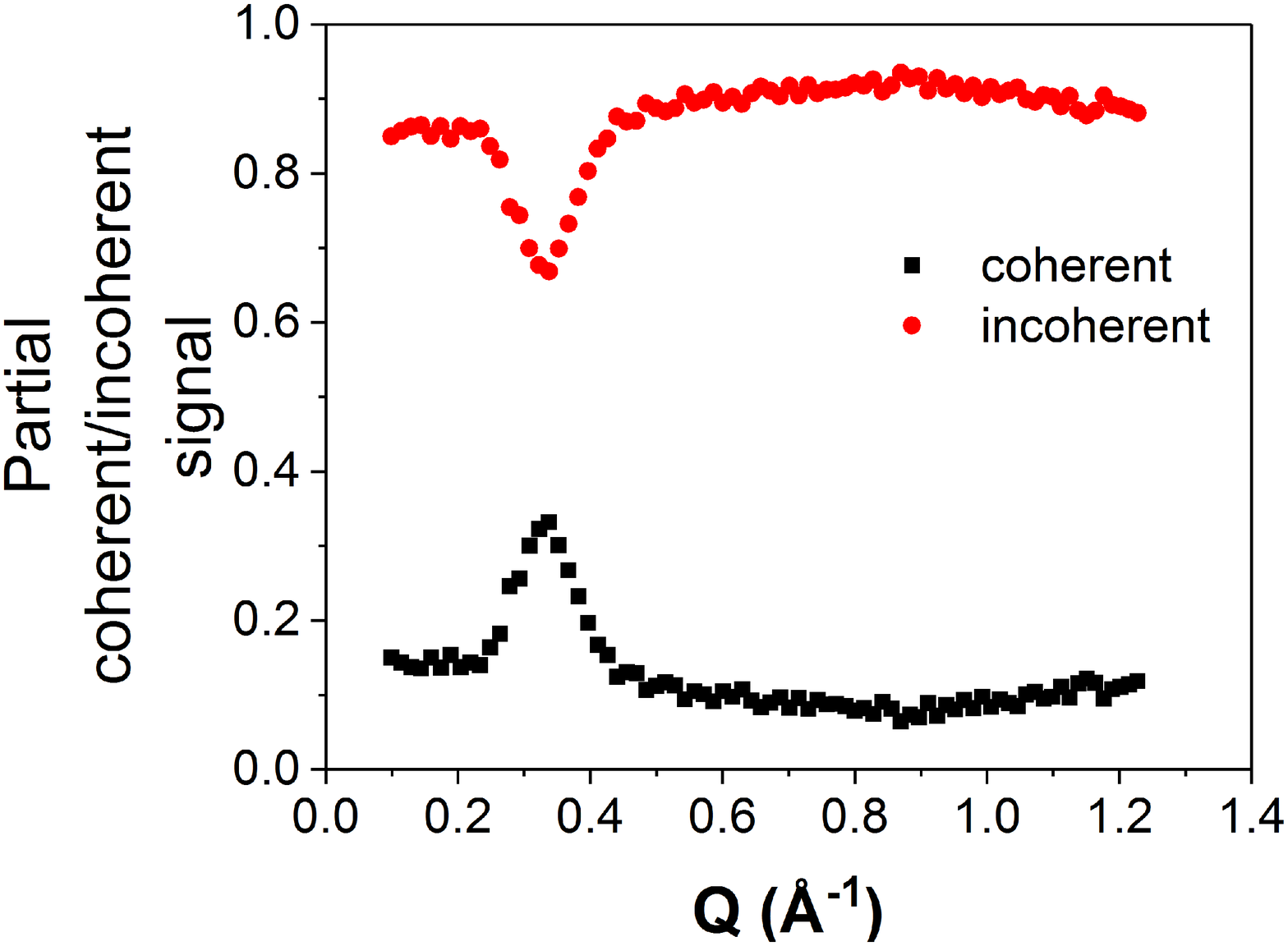}
\caption{}\label{fig:spin_echo_coh_inc}
\end{subfigure}
\caption{(a) QENS signal of h-RR-P3HT. h-RRa-P3HT, d-RR-P3HT and d-RRa-P3HT at 296K, from the time-of-flight measurements using IN6, illustrating the difference in signal amplitude for coherent (deuterated) and incoherent (hydrogenated) samples. (b) Spin-echo coherent/incoherent signal ratio for h-RR-P3HT highlighting the dominant character of the incoherent signal, against the coherent one, stemming from neutron scattering of hydrogenated samples.} \label{fig:coh_inc}
\end{figure}

Dynamics of polymeric systems spans a large time scale. Therefore in the present study we combine Quasi-Elastic Neutron Scattering (QENS) and Neutron Spin-Echo (NSE) to cover both the picosecond and nanosecond time domains. As neutron based techniques, and in addition to the unique feature allowing to use deuteration for a contrast variation purpose against other techniques of probe, QENS and NSE offer the possibility to cover suitably the dynamical space in terms of Q-resolution as well as the energy resolution, in comparison for instance with dielectric spectroscopy or solid-state nuclear magnetic resonance. Thus, enabling to correlate efficiently structure (Q-resolution) and dynamics (energy resolution).\\
Slow dynamics of polymers such as rotation and reorientation of the side chains, and backbone torsion, can be expressed in terms of the density correlation function, which depends on both space and time:
\begin{equation}
    \rho G(\mathbf{r}, \mathbf{r'}, t) = <\rho(\mathbf{r}+\mathbf{r'},t)\rho(\mathbf{r'},0)>
    \label{eq:density_fluctuation}
\end{equation}
For a homogeneous system, equation \ref{eq:density_fluctuation} can be rewritten as:
\begin{equation}
\begin{aligned}
    G(\mathbf{r}, t) & = \frac{1}{N}<\sum_{i=0}^n \sum_{j=0}^n \delta(\mathbf{r}-\mathbf{r_i}(t)+\mathbf{r_j(0)}) \\
    & = \frac{1}{N}<\sum_{i=0}^n \delta(\mathbf{r}-\mathbf{r_i}(t)+\mathbf{r_i(0)}>
    + \frac{1}{N}<\sum_{i=0}^n \sum_{j\neq i}^n \delta(\mathbf{r}-\mathbf{r_i}(t)+\mathbf{r_j(0)}> \\
    & = G_s(\mathbf{r},t) + G_d(\mathbf{r},t)
\end{aligned}
    \label{eq:van_hove_splitting}
\end{equation}
Where $G(\mathbf{r}, t)$ is the Van Hove function. It can be expressed in terms of two terms; a self-part contribution, denoted $G_s(\mathbf{r}, t)$, and a distinct part contribution, denoted $G_d(\mathbf{r}, t)$. $G_s(\mathbf{r}, t)$ is the  probability density of finding a particle i at a position $\mathbf{r}$, at a time t, knowing that this particle was in the same position at the reference time t=0. $G_d(\mathbf{r}, t)$ is the probability density of finding a particle i at a position $\mathbf{r}$, at time t knowing that a second particle j was at $\mathbf{r}$, at the origin t=0. $G_d(\mathbf{r}, 0)$ is proportional to the pair distribution function. $G_s(\mathbf{r}, t)$ is related to the self-motions while $G_d(\mathbf{r}, t)$ is related to the collective motions. Since neutron measurements are carried out in the reciprocal space, it is therefore of interest to introduce the intermediate scattering function, expressed in the time domain, $F(\mathbf{Q},t)$:
\begin{equation}
    F(\mathbf{Q},t) = \int d\mathbf{r} G(\mathbf{r}, t)e^{-i\mathbf{Q}.\mathbf{r}} = F_s(\mathbf{Q},t) + F_d(\mathbf{Q},t)
    \label{eq:FFT_to_intermediate}
\end{equation}
The physical interest of writing down the intermediate scattering function, $F(\mathbf{Q},t)$, in terms of the self part and the distinct part, is that samples with a larger and dominant incoherent neutron cross-section than coherent cross-section, i.e. hydrogenated samples, give access to the self part of the intermediate scattering function. While deuterated samples lead to a comparable coherent and incoherent contribution of the self part and thus give access to the total intermediate scattering function, $F(\mathbf{Q},t)$. Therefore, knowing $F_d(\mathbf{Q},t)$ requires a knowledge of $F(\mathbf{Q},t)$ and $F_s(\mathbf{Q},t)$.\\ 
Intermediate scattering function can be directly accessed by NSE technique, or indirectly via QENS measurements. QENS provides a direct measurement of the dynamical structure factor $S(\mathbf{Q}, \omega)$, convoluted with an instrumental resolution $R(\mathbf{Q}, \omega)$. $S(\mathbf{Q}, \omega)$ and $F(\mathbf{Q},t)$ are related to each other by a FT, as:
\begin{equation}
    S(\mathbf{Q},\omega) = \int dt F(\mathbf{Q}, t)e^{-i\omega t} 
\end{equation}
In the following, we present neutron scattering measurements using (i) the time-of-flight spectrometer IN6, whose time window covers the picosecond time-scale, (ii) the backscattering spectrometer IN16B, to reach the nanosecond time-scale, and (iii) the NSE IN11 allowing a direct probe of the relaxation in the time domain, from the picosecond to the nanosecond time-scale. QENS is primarily used with hydrogenated samples as the incoherent signal is much stronger than the coherent signal, thus also relevant to better optimize the acquisition time (Figure \ref{fig:QENS_coh_inc}). However, since in hydrogenated samples the scattering is dominated by the incoherent scattering (Figure \ref{fig:spin_echo_coh_inc}), we went a step further by extending our neutron measurements to deuterated samples, in order to also probe the collective motions of P3HT. 

\begin{figure}[H]
\begin{subfigure}[b]{0.45\textwidth}
\includegraphics[width=\textwidth]{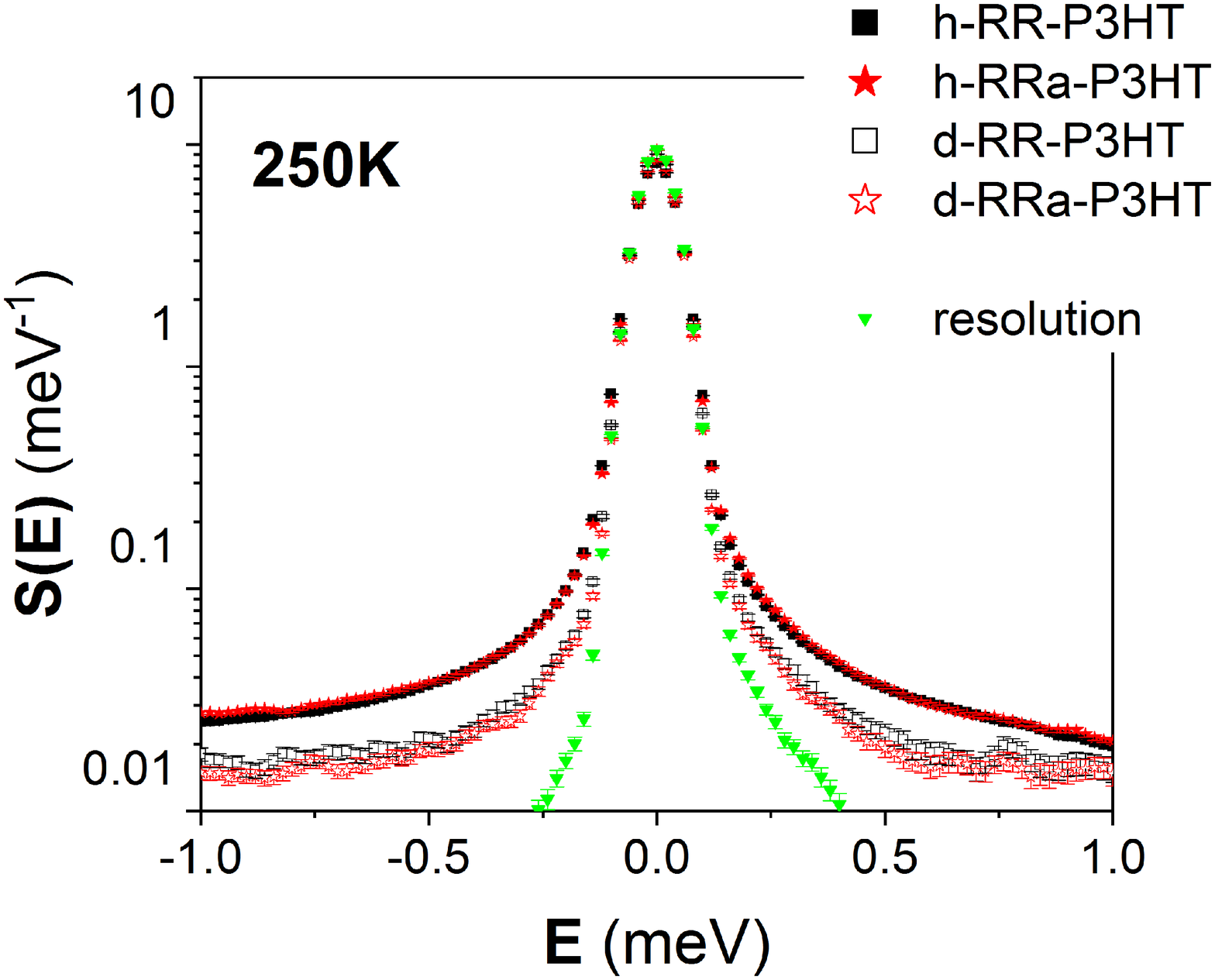}
\caption{}\label{fig:QENS_IN6_250K}
\end{subfigure}	
\begin{subfigure}[b]{0.45\textwidth}
\includegraphics[width=\textwidth]{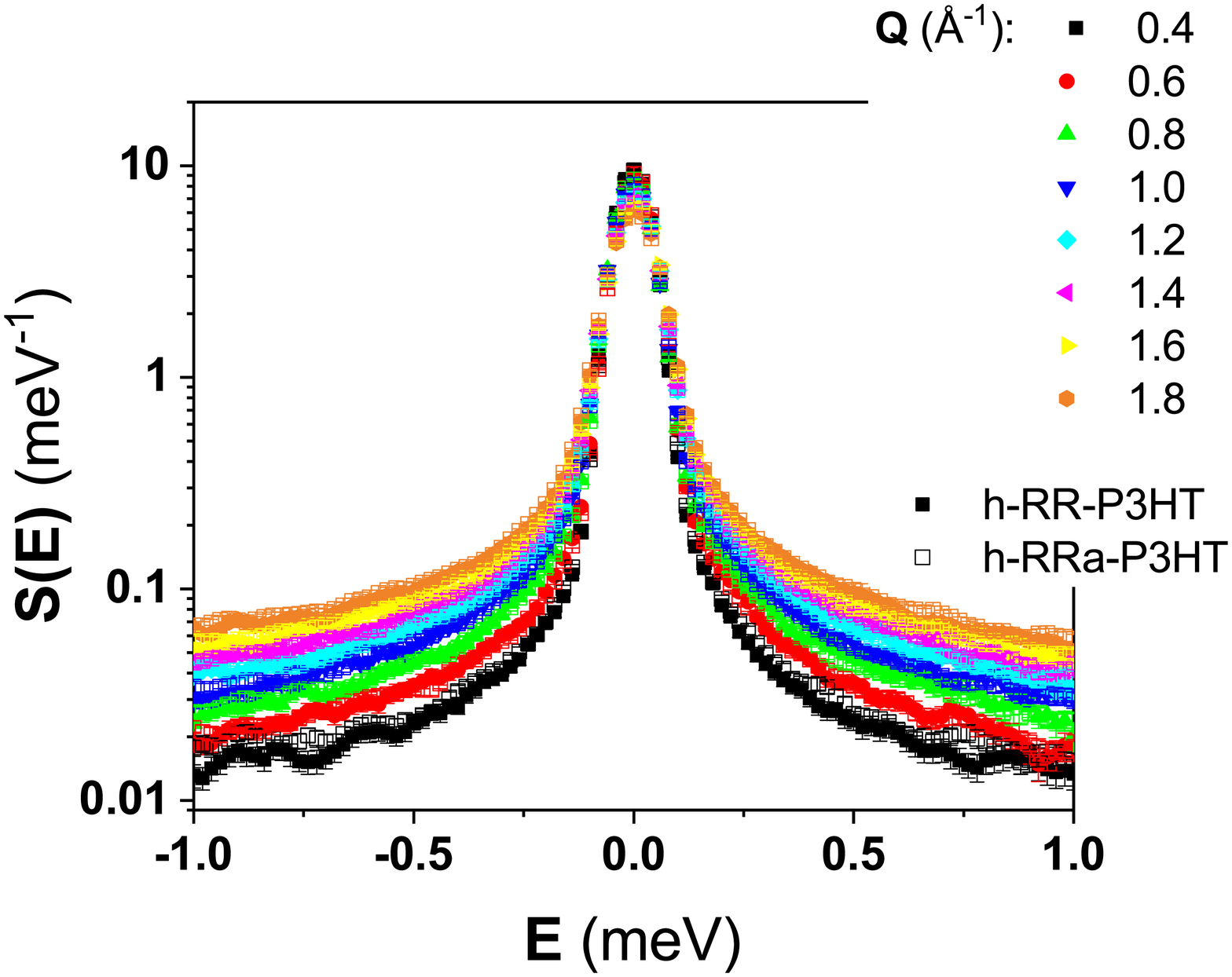}
\caption{}\label{fig:QENS_h_RR_RRa}
\end{subfigure}
\begin{subfigure}[b]{0.45\textwidth}
\includegraphics[width=\textwidth]{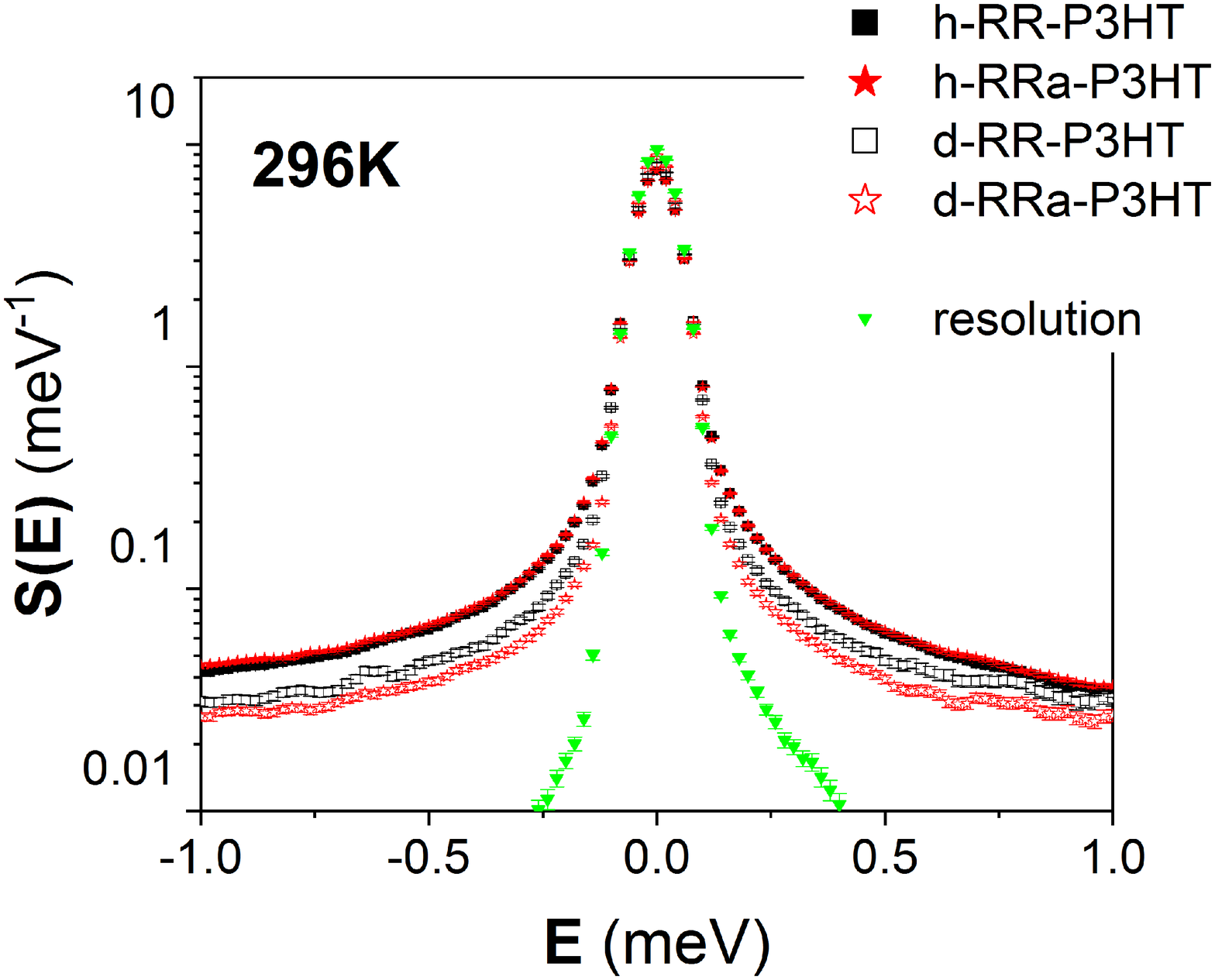}
\caption{}\label{fig:QENS_IN6_296K}
\end{subfigure}	
\begin{subfigure}[b]{0.45\textwidth}
\includegraphics[width=\textwidth]{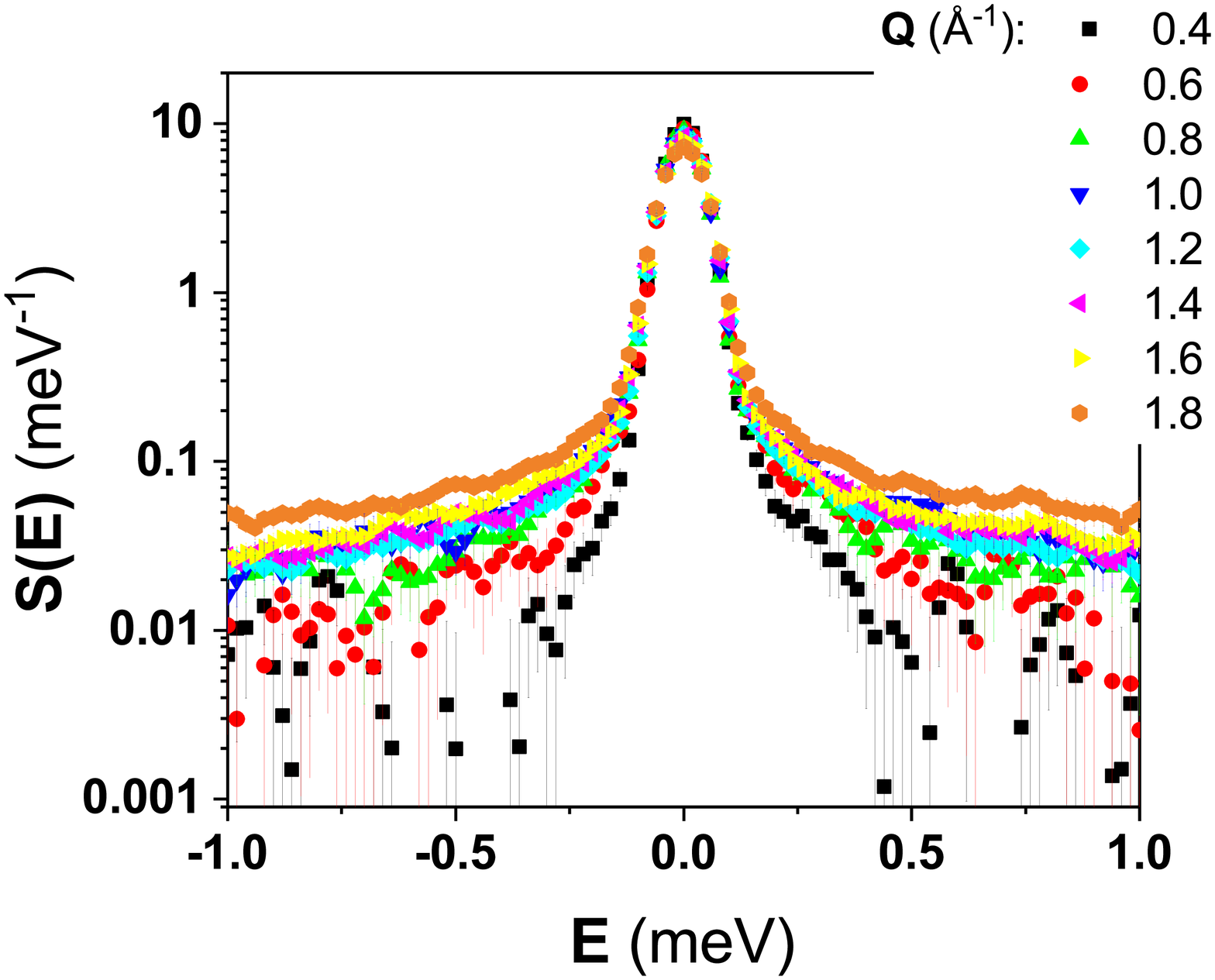}
\caption{}\label{fig:QENS_d-RR}
\end{subfigure}
\begin{subfigure}[b]{0.45\textwidth}
\includegraphics[width=\textwidth]{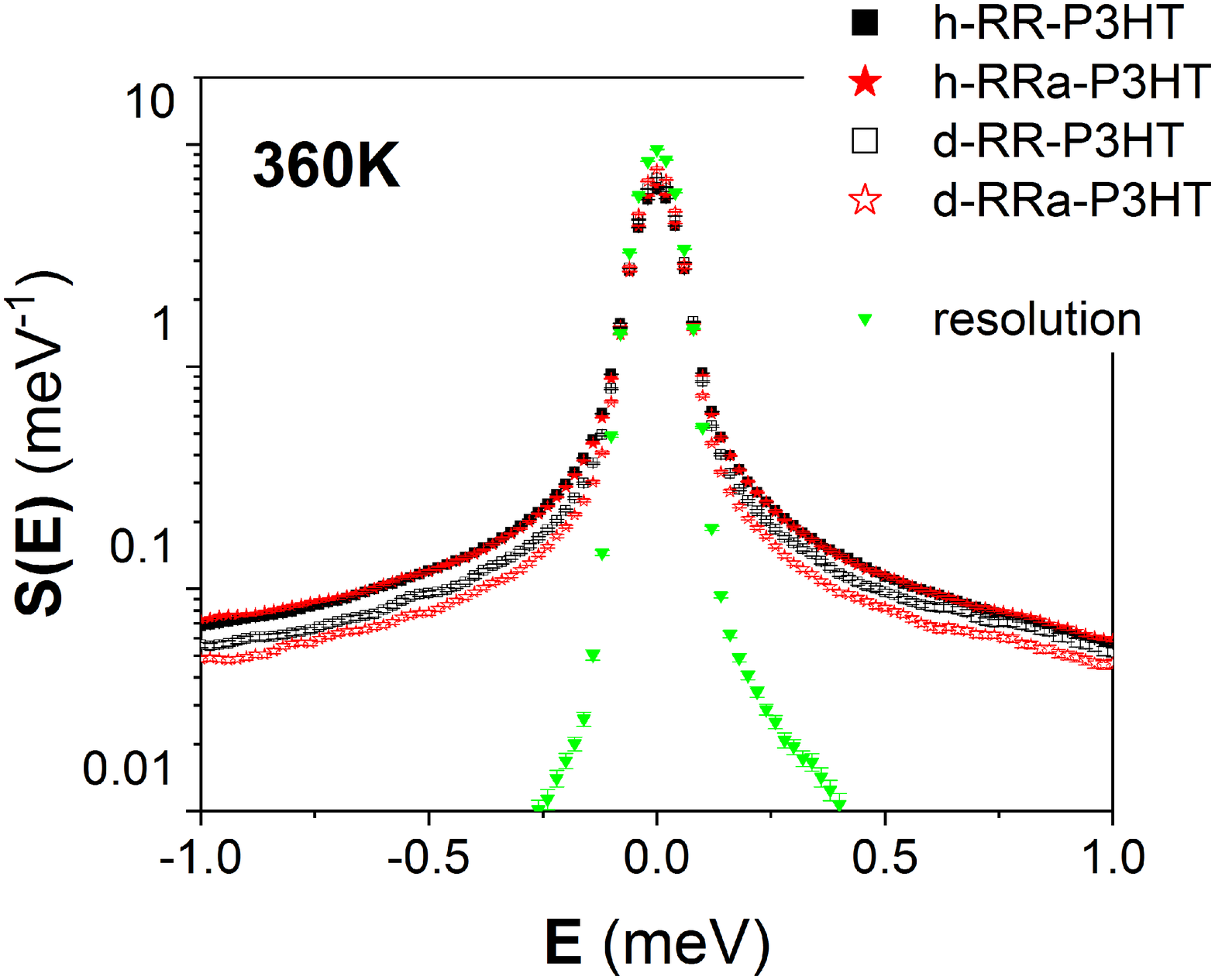}
\caption{}\label{fig:QENS_IN6_360K}
\end{subfigure}
\begin{subfigure}[b]{0.45\textwidth}
\includegraphics[width=\textwidth]{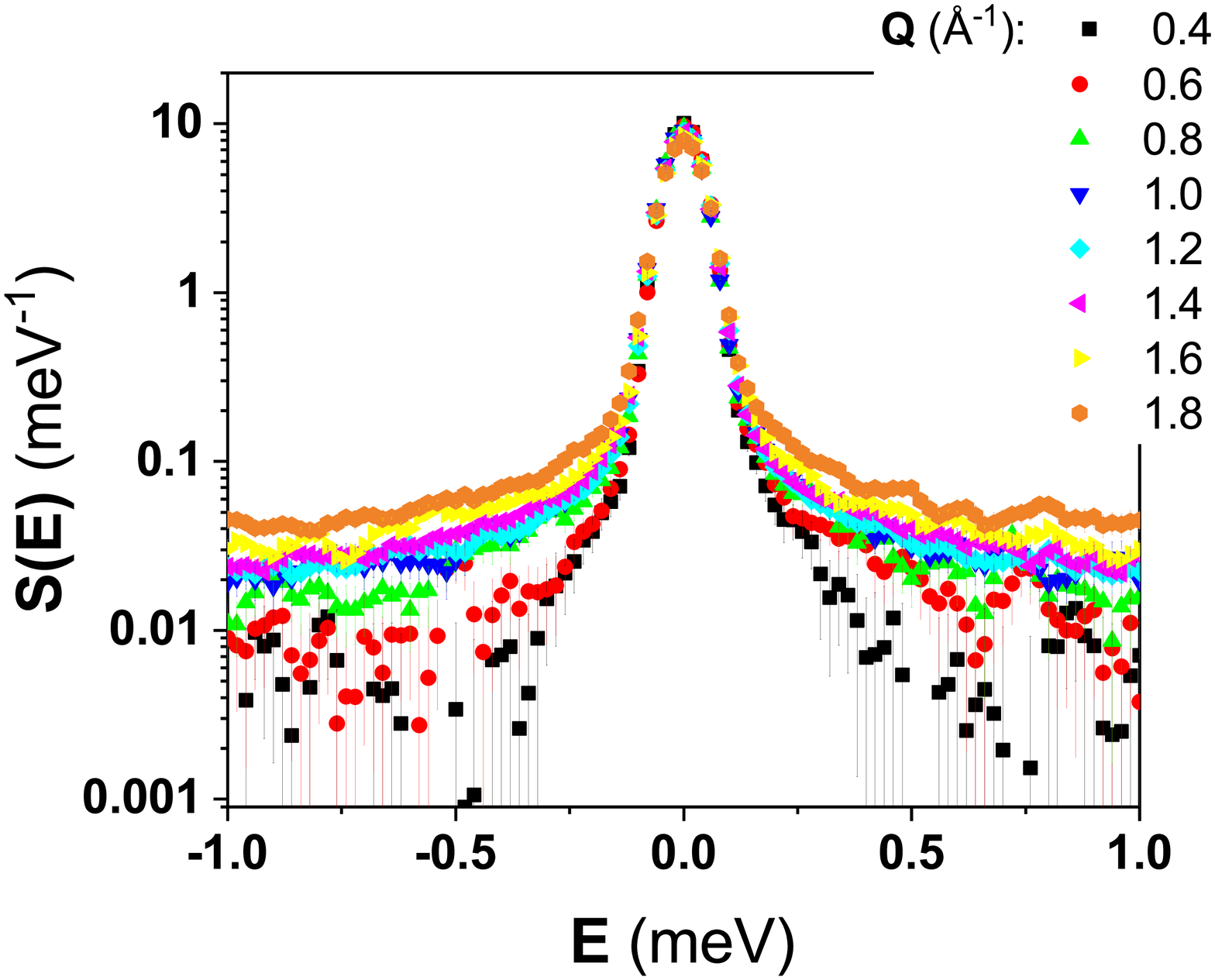}
\caption{}\label{fig:QENS_d-RRa}
\end{subfigure}
\caption{Left panels: the Q-averaged TOF-based QENS spectra from IN6 measurements, for h-RR-P3HT, h-RRa-P3HT, d-RR-P3HT and d-RRa-P3HT at (a) 250K, (c) 296K and (e) 360K. The instrumental resolution was determined from a vanadium sample at 296K. Right panels: Q-dependent QENS spectra of (b) h-RR-P3HT and h-RRa-P3HT, (d) d-RR-P3HT and (f) d-RRa-P3HT at 296K. Note that h-RR-P3HT and h-RRa-P3HT exhibit nearly the same behavior, within the error bars.} \label{fig:QENS_IN6}
\end{figure}

\begin{figure}[H]
\begin{subfigure}[b]{0.44\textwidth}
\includegraphics[width=\textwidth]{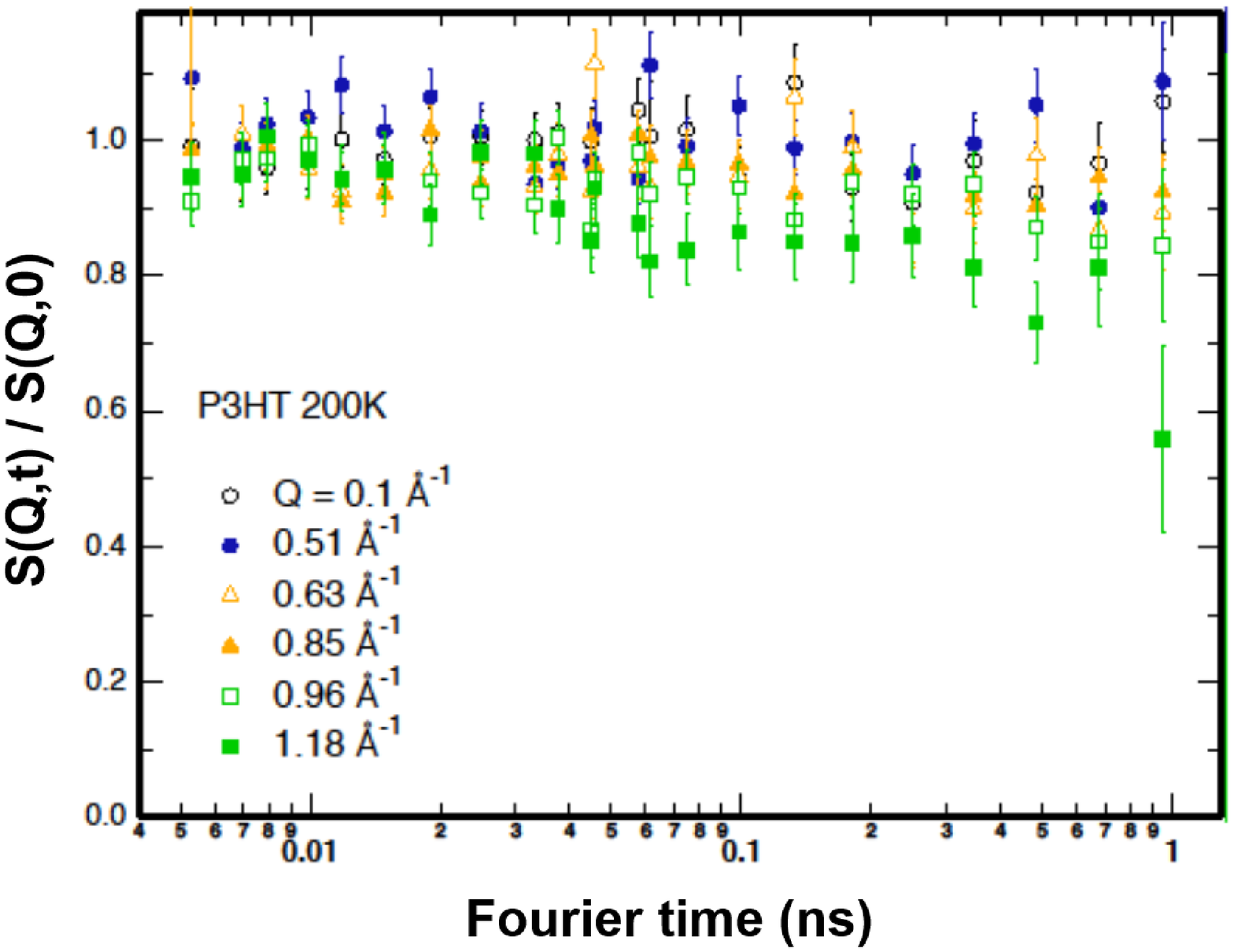}
\caption{}\label{fig:intermediate_200K}
\end{subfigure}
\begin{subfigure}[b]{0.44\textwidth}
\includegraphics[width=\textwidth]{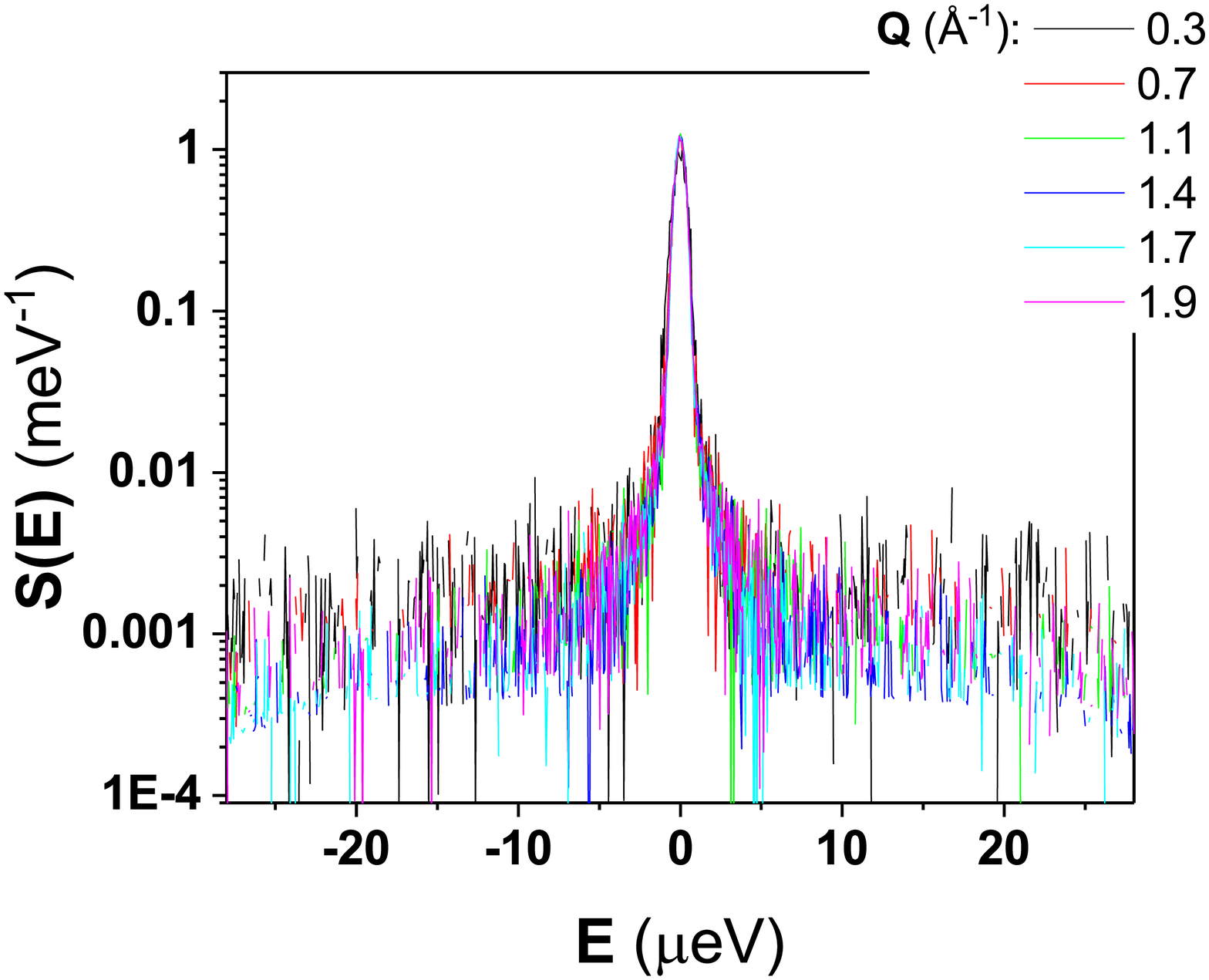}
\caption{}\label{fig:IN16B_180K}
\end{subfigure}
\begin{subfigure}[b]{0.44\textwidth}
\includegraphics[width=\textwidth]{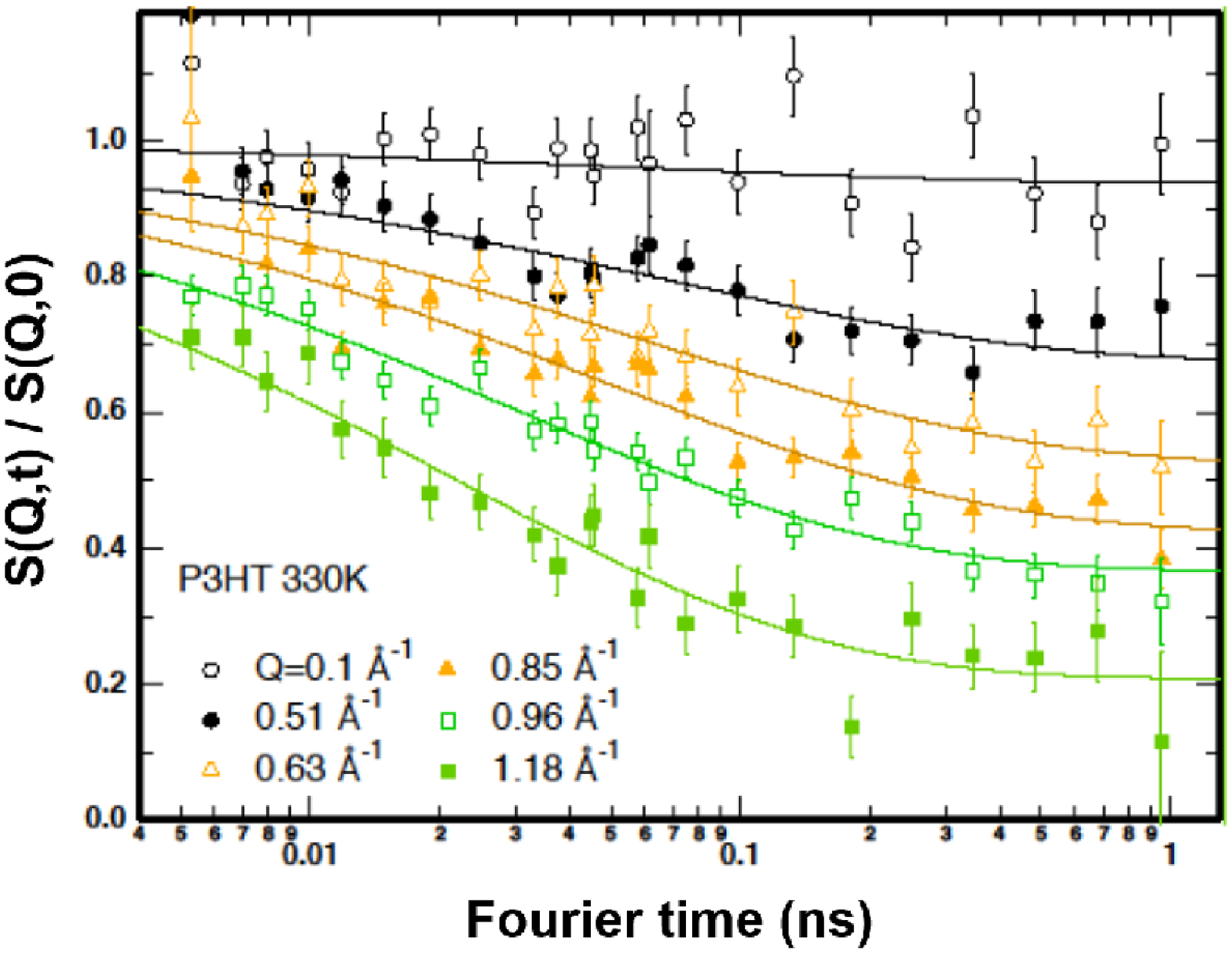}
\caption{}\label{fig:intermediate_330K}
\end{subfigure}
\begin{subfigure}[b]{0.44\textwidth}
\includegraphics[width=\textwidth]{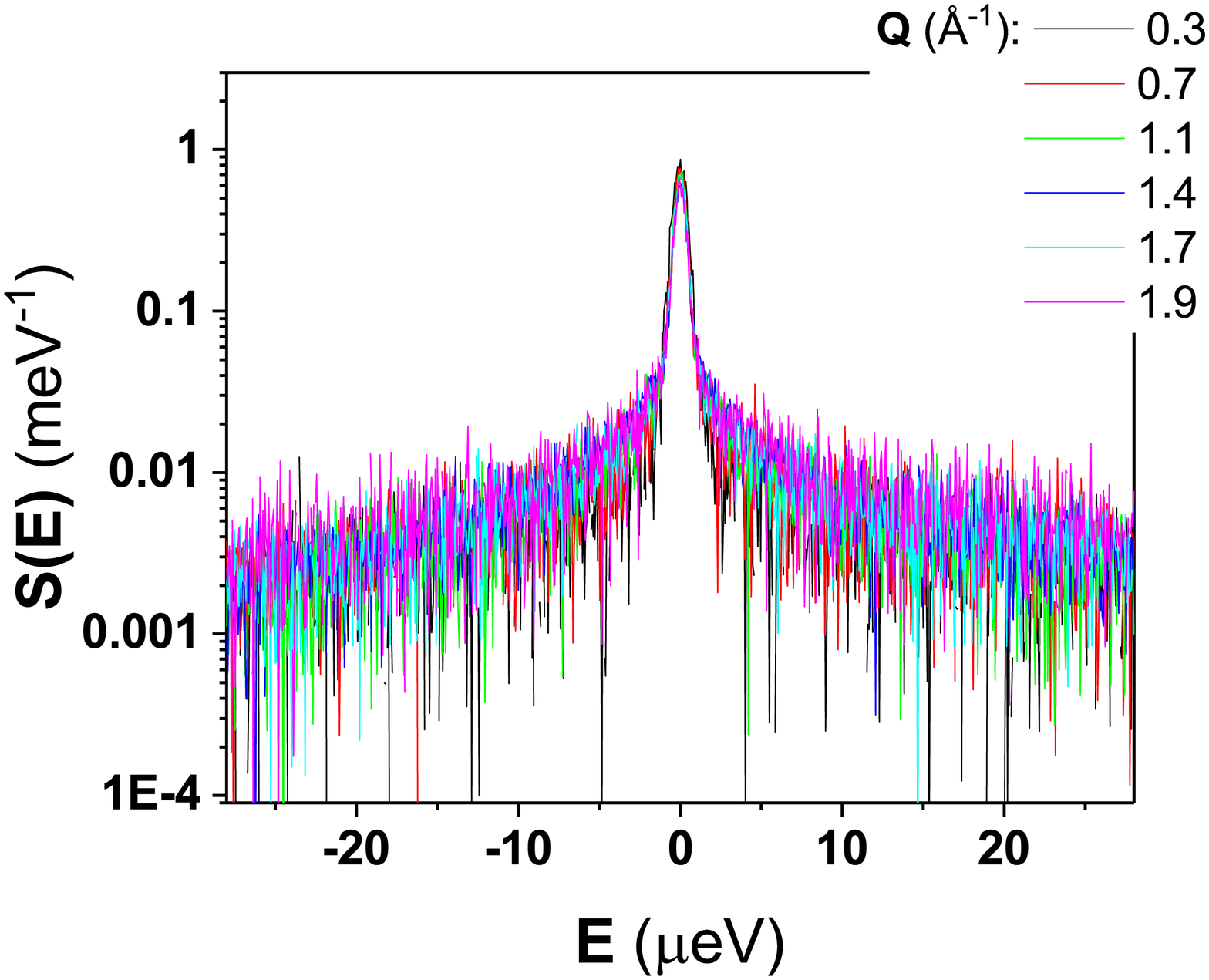}
\caption{}\label{fig:IN16B_360K}
\end{subfigure}
\caption{Left panels: intermediate scattering function of h-RR-P3HT from spin-echo measurement using IN11, at (a) 200K and (c) 330K. Right panels: the Q-dependent QENS spectra of d-RR-P3HT measured on IN16B at (b) 180K and (d) 360K. }\label{fig:Intermediate_scattering_function}
\end{figure}

\begin{figure}[H]
\begin{subfigure}[b]{0.45\textwidth}
\includegraphics[width=\textwidth]{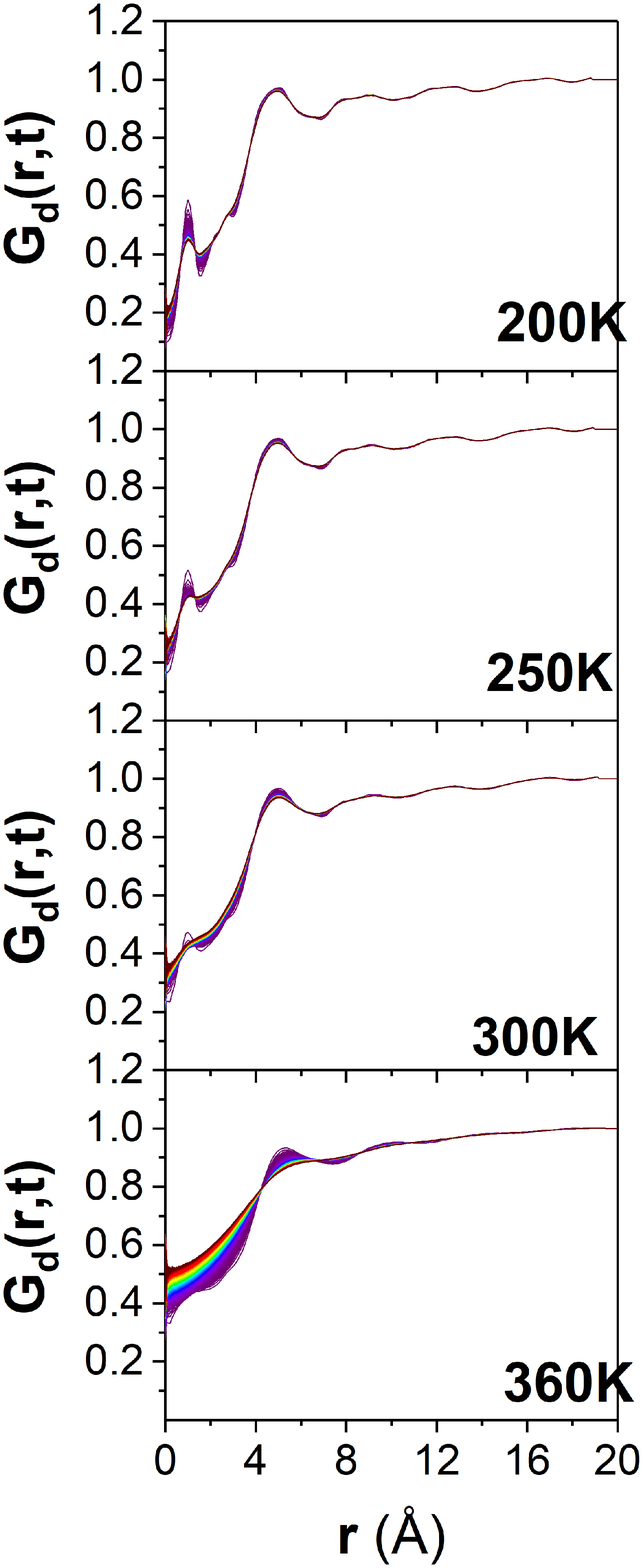}
\caption{}\label{fig:distinct}
\end{subfigure}
\begin{subfigure}[b]{0.45\textwidth}
\includegraphics[width=\textwidth]{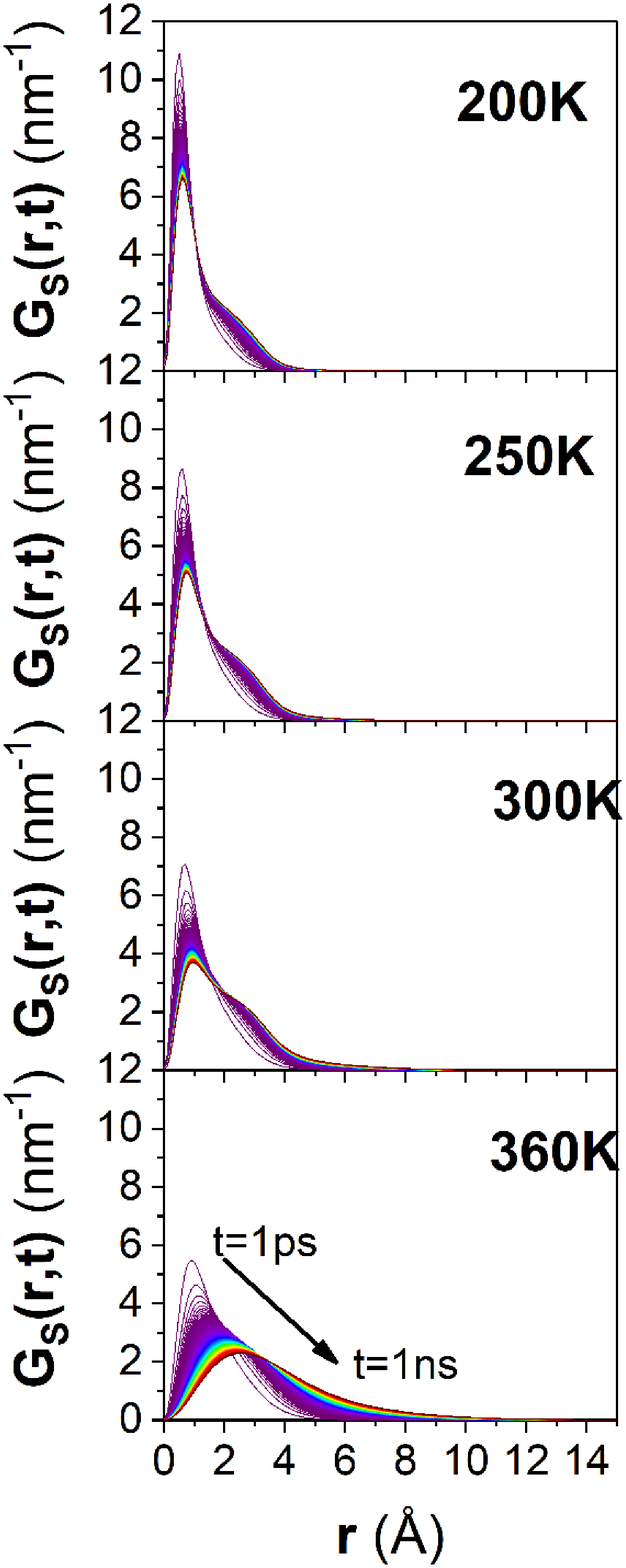}
\caption{}\label{fig:self_part}
\end{subfigure}
\caption{Time-evolution of the neutron-weighted (a) distinct G$_d$(r,t) and (b) self-part G$_s$(r,t) of the Van Hove function as a function of temperature for crystalline (a) d-RR-P3HT (coherent scattering dominating) and (b) h-RR-P3HT (incoherent scattering dominating).}\label{fig:Van_Hove}	
\end{figure}

\begin{figure}[H]
\begin{subfigure}[b]{0.27\textwidth}
\includegraphics[width=\textwidth]{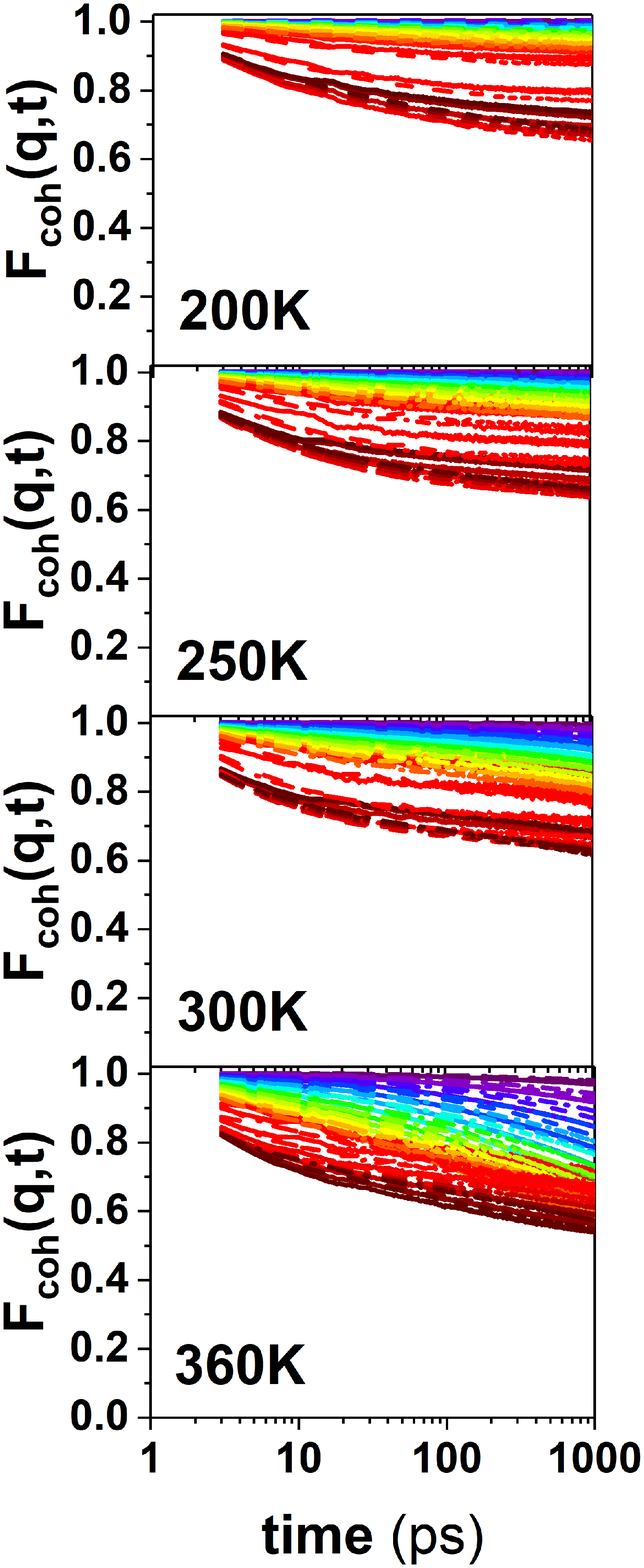}
\caption{}\label{fig:coherent_intermediate}
\end{subfigure}
\begin{subfigure}[b]{0.45\textwidth}
\includegraphics[width=\textwidth]{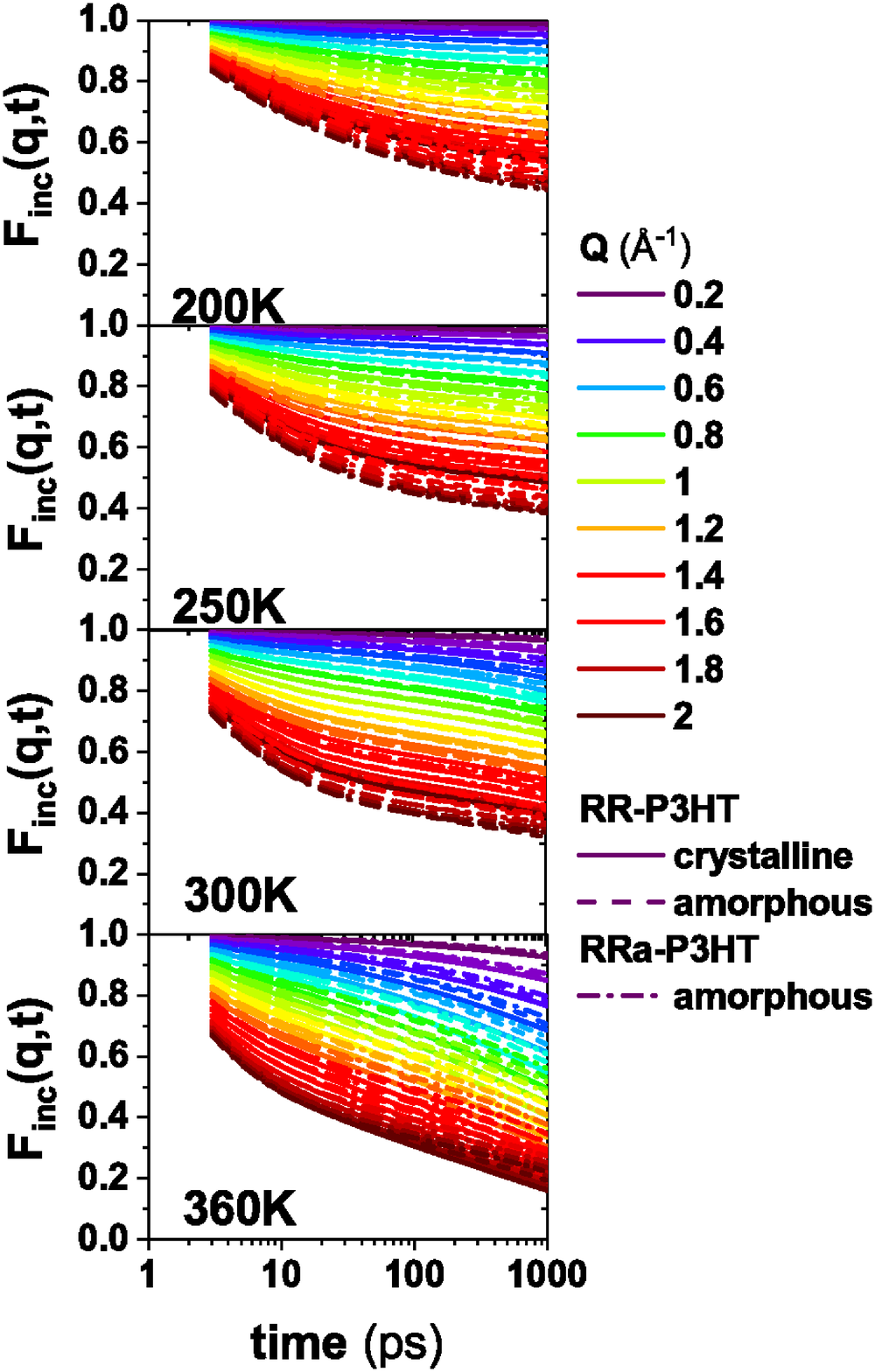}
\caption{}\label{fig:incoherent_intermediate}
\end{subfigure}
\caption{The coherent $F_{coh}(q,t)$ (a), and incoherent $F_{inc}(q,t)$ (b) intermediate scattering function as a function of temperature and Q, calculated from MD simulations for the different studied models.}\label{fig:Intermediate_scattering_function_MD}	
\end{figure}

\begin{figure}[H]
\begin{subfigure}[b]{0.45\textwidth}
\includegraphics[width=\textwidth]{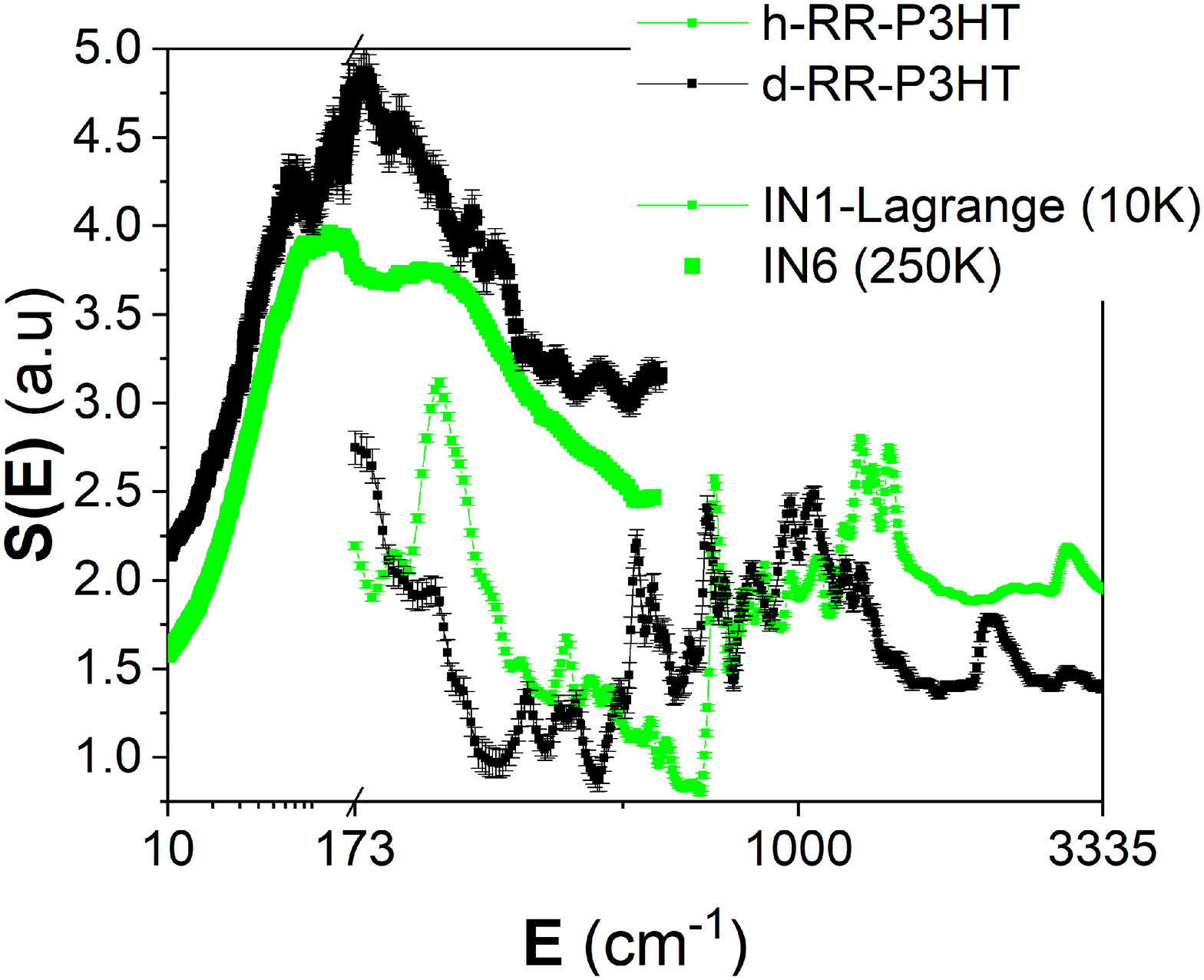}
\caption{}\label{fig:IN1_a}
\end{subfigure}
\begin{subfigure}[b]{0.45\textwidth}
\includegraphics[width=\textwidth]{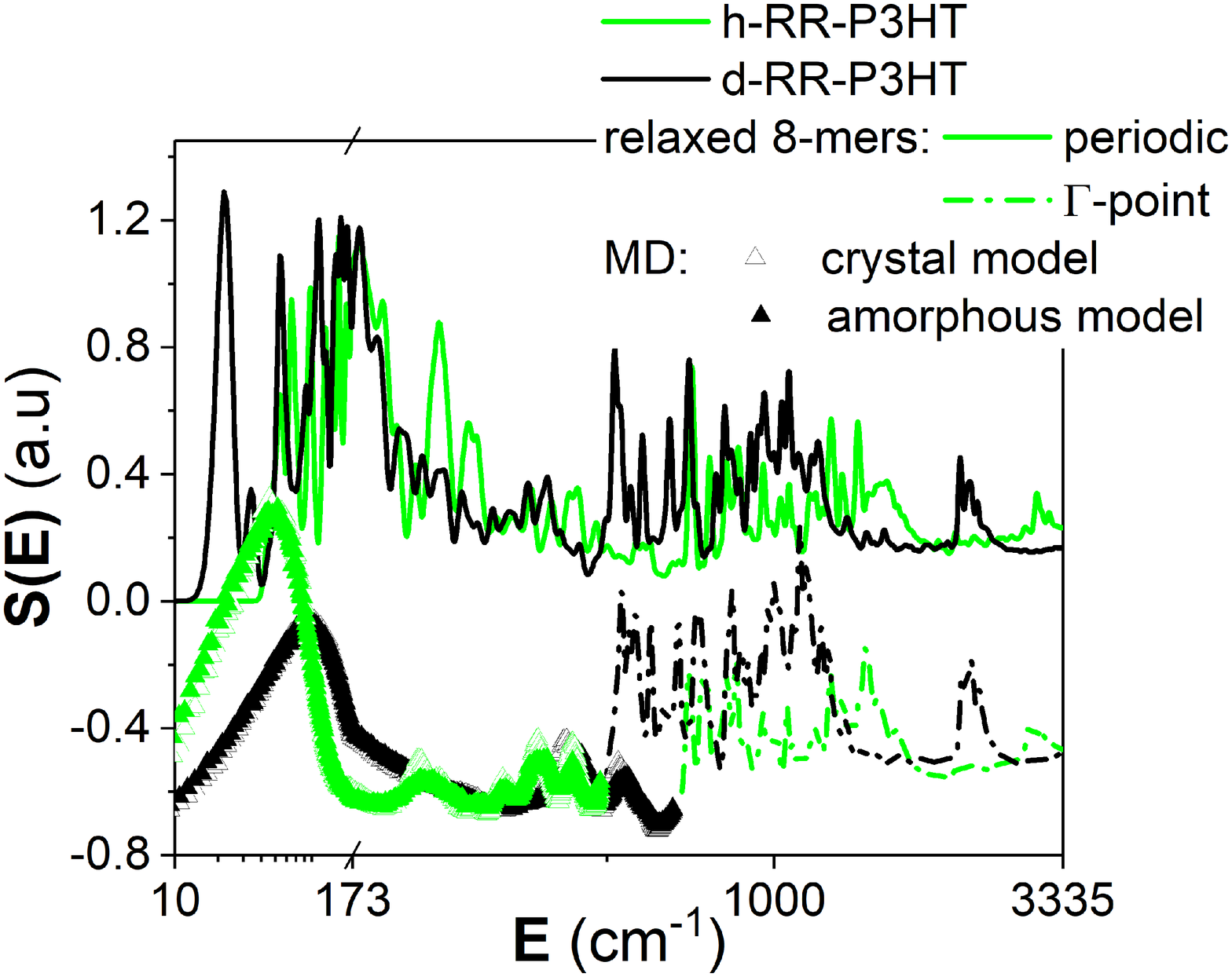}
\caption{}\label{fig:IN1_b}
\end{subfigure}
\begin{subfigure}[b]{0.45\textwidth}
\includegraphics[width=\textwidth]{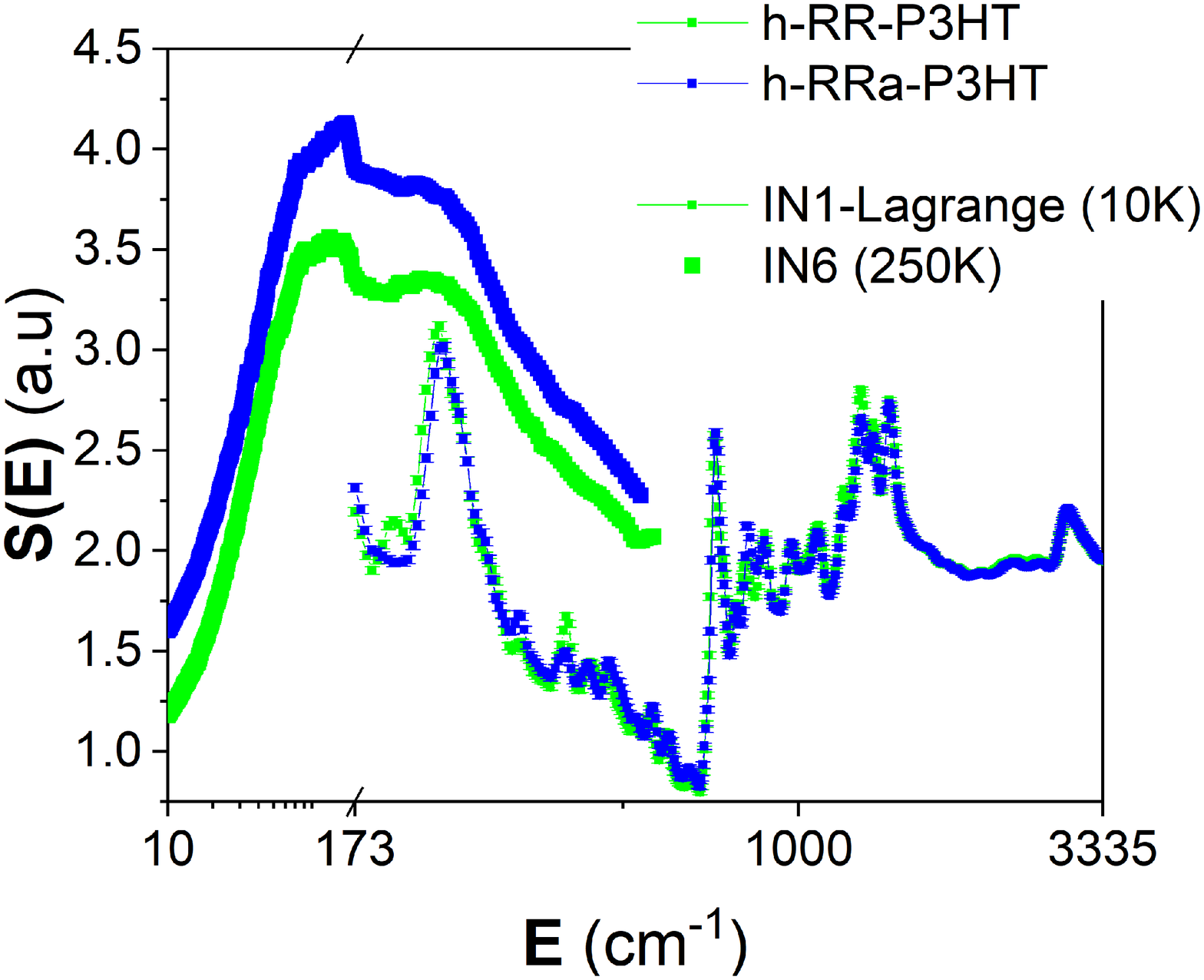}
\caption{}\label{fig:IN1_c}
\end{subfigure}
\begin{subfigure}[b]{0.45\textwidth}
\includegraphics[width=\textwidth]{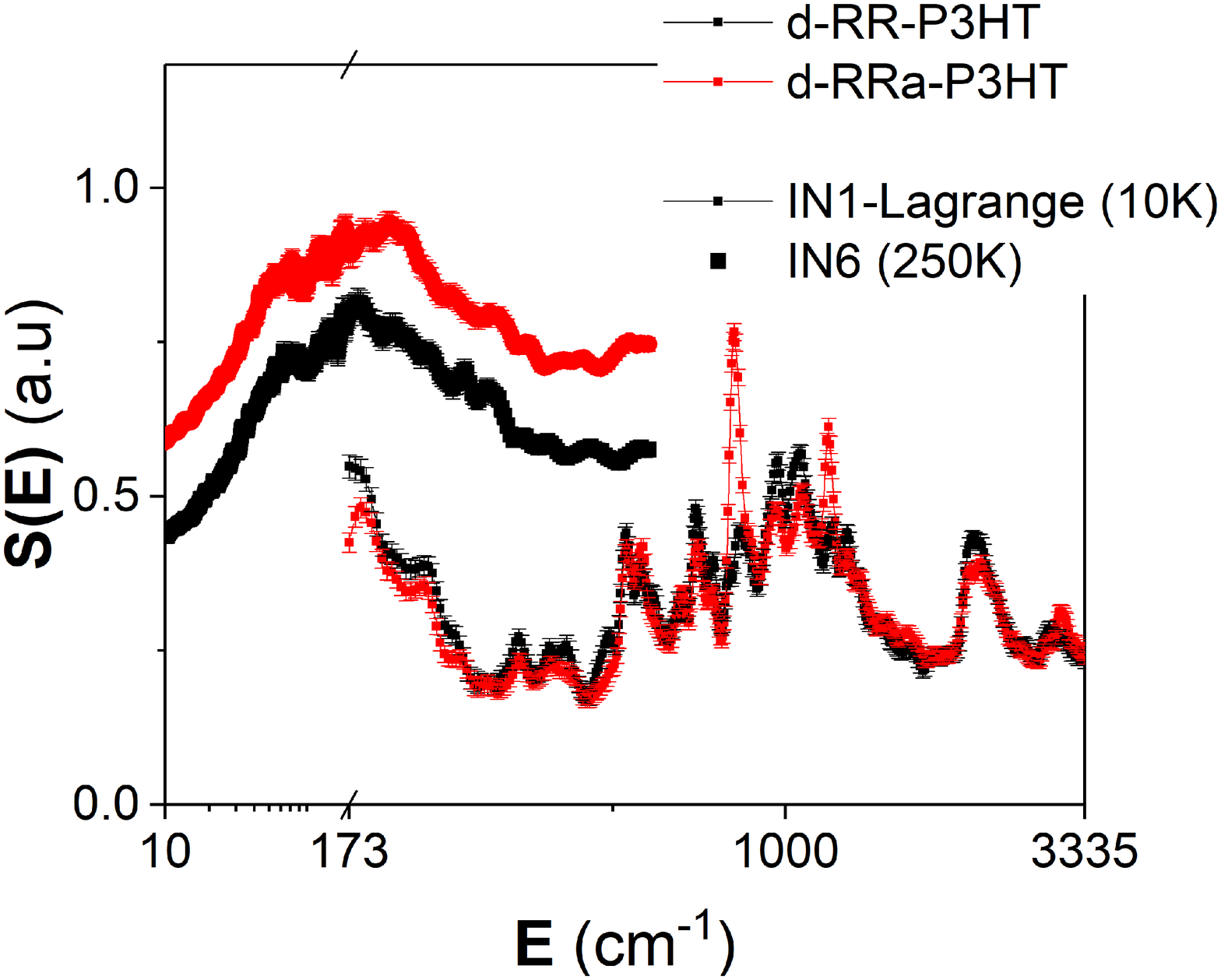}
\caption{}\label{fig:IN1_d}
\end{subfigure}
\begin{subfigure}[b]{0.45\textwidth}
\includegraphics[width=\textwidth]{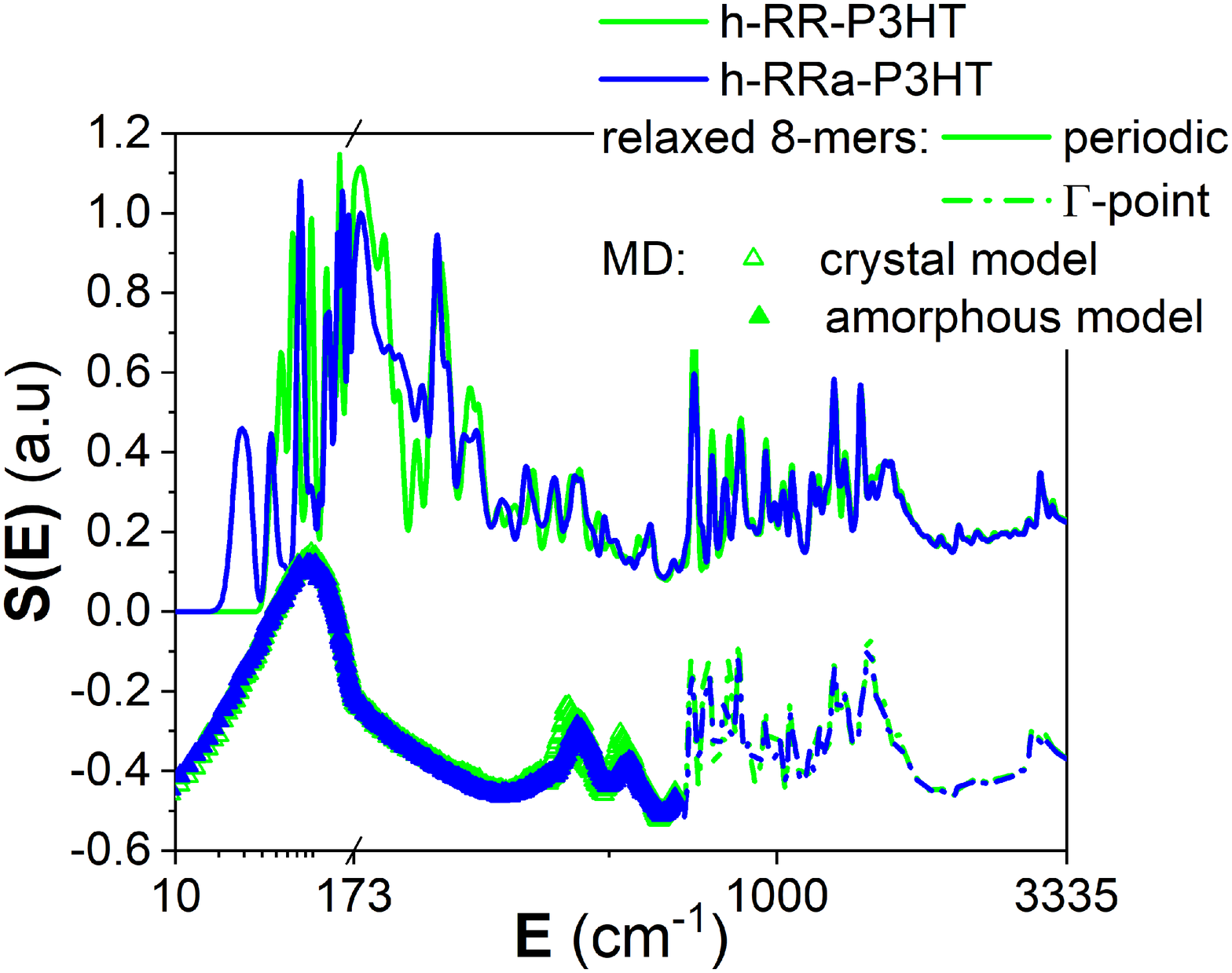}
\caption{}\label{fig:INS_h_calculated}
\end{subfigure}
\begin{subfigure}[b]{0.45\textwidth}
\includegraphics[width=\textwidth]{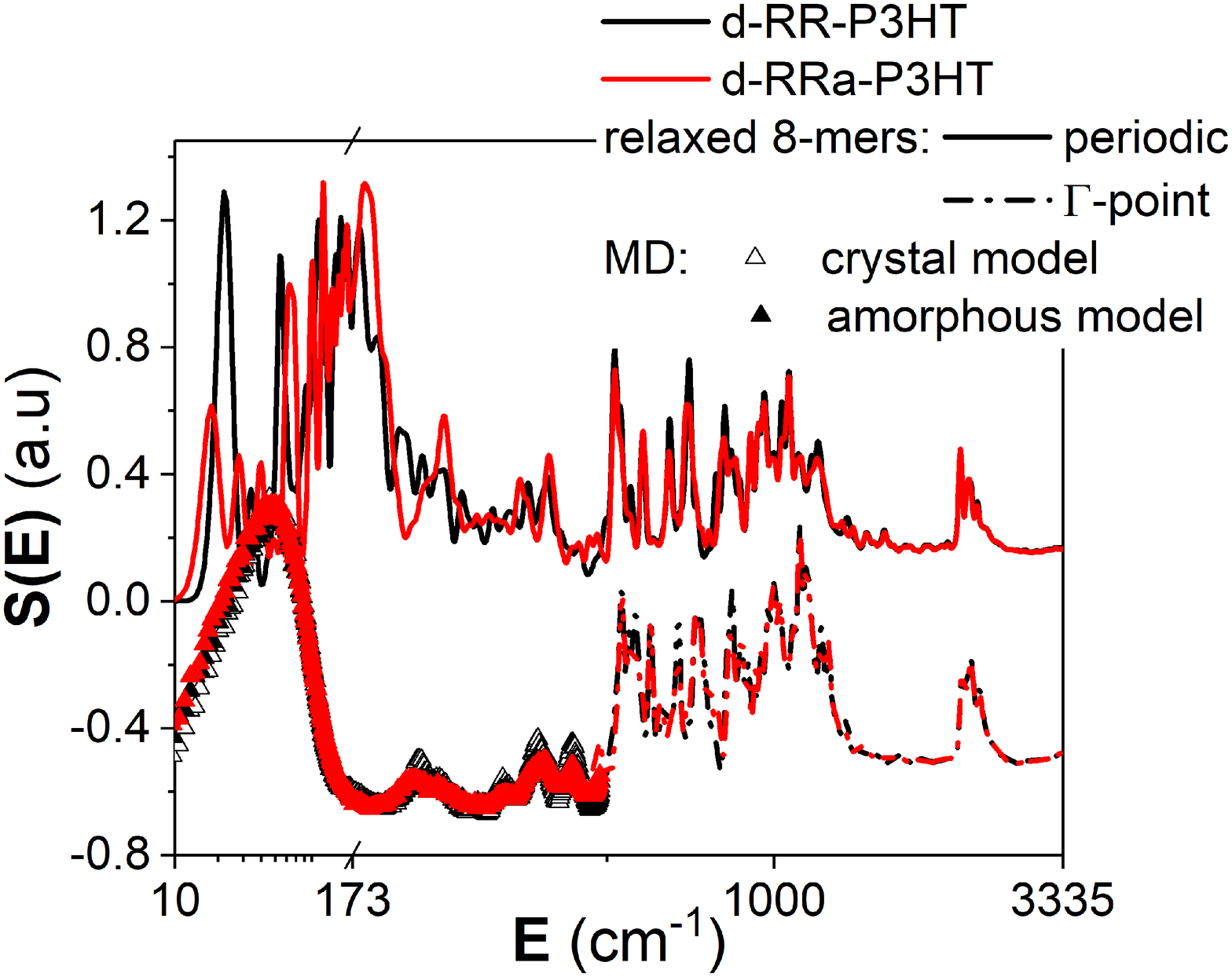}
\caption{}\label{fig:INS_d_calculated}
\end{subfigure}
\caption{INS spectra measured using IN6, at 250K, and IN1-Lagrange, at 10K of (a) h-RR-P3HT and d-RR-P3HT, (c) h-RR-P3HT and h-RRa-P3HT and, (d) d-RR-P3HT and d-RRa-P3HT. Calculated INS spectra from relaxed 8-mers model using DFT-based periodic and isolated molecular approaches, as well as classical MD simulations performed on crystalline and amorphous boxes of (b) h-RR-P3HT and d-RR-P3HT, (e) h-RR-P3HT and h-RRa-P3HT and, (f) d-RR-P3HT and d-RRa-P3HT. The presentation of the energy range is restricted to show the low-energy external lattice modes, in the case of the MD simulation, followed by the high-energy range of the internal molecular modes, in the case of the molecular simulations.}\label{fig:IN1}
\end{figure}

The TOF measurements on IN6 were performed using an incident neutron wavelength of 5.12 \r{A}, leading to an elastic resolution of about 70 $\mu$eV, as determined by a vanadium standard, which allows to probe dynamics with relaxation times of the order of the picosecond. Figures \ref{fig:QENS_IN6_250K}, \ref{fig:QENS_IN6_296K} and \ref{fig:QENS_IN6_360K} present the outcome of the TOF-based QENS measurements using IN6.\\
The temperature range 250K-360K was targeted for the measurements since it covers the glass transition of the P3HT polymers. No differences in dynamics is observed between h-RR-P3HT and h-RRa-P3HT. The strong incoherent scattering of the hydrogen atoms located mainly on the side chains dominates the QENS signal. Thus, on the measured time scale, the self-motion of the hydrogens is not impacted by the difference in microstructure (semi-crystallinity or backbone conformation). However, d-RR-P3HT spectra appears broader than d-RRa-P3HT, pointing towards a faster dynamics. Note that it is not straightforward to interpret deuterated QENS signals, as the purely elastic line is impacted by the Bragg peaks. Within the temperature range 250-360K, the hydrogenated samples exhibit broader peaks, reflecting a faster dynamics, within the accessible picosecond time window of the IN6 measurements, than the deuterated samples, although at higher temperature, this difference is reduced (Figure \ref{fig:QENS_IN6_250K}, \ref{fig:QENS_IN6_296K} and \ref{fig:QENS_IN6_360K}). It is important to note that different dynamics are probed. In the case of the hydrogenated samples, the self-motions of the hydrogen atoms is dominant and therefore mostly measured, while in the case of the deuterated samples, both the self-motions of all the atoms - as deuterium has a incoherent neutron cross-section close to carbon and sulfur - as well as the collective motions are measured. In the hydrogenated samples, a clear increase of the broadening of the elastic peak is observed as Q increases (Figure \ref{fig:QENS_h_RR_RRa}). Note that h-RR-P3HT and h-RRa-P3HT exhibit nearly the same behavior, within the error bars. For the deuterated samples, the Q-dependence is less pronounced, especially within the Q-range 1.0-1.4 \r{A}$^{-1}$.\\
The QENS spectra of the hydrogenated samples were fitted using the following model:
\begin{equation}
S'(\mathbf{Q},\omega) = R(\mathbf{Q}, \omega) \otimes \{A_0(\mathbf{Q})\delta(\omega)+A_1(\mathbf{Q})L(\mathbf{Q},\omega)+background(\mathbf{Q})\}
\label{eq:fit}
\end{equation}
where $A_0(\mathbf{Q})$ is the intensity of the elastic peak, $L(\mathbf{Q},\omega)$ is a Lorentzian of intensity $A_1(\mathbf{Q}$) and width $\Gamma(\mathbf{Q})$, describing the dynamics captured by the spectrometer, and the background describes the dynamics that are too fast for the instrumental time window. The resolution of the spectrometer, $R(\mathbf{Q}, \omega)$, was determined using a vanadium standard at 296K.\\
The elastic incoherent structure factor (EISF) (Figure S11 in the SI), $A_0$, decreases as a function of Q, and this behavior is more pronounced with temperature. Both the contribution of the Lorentzian and the background increase with temperature and Q (Figure S11 in the SI). More dynamics are captured within the energy/time window of IN6 as the momentum transfer, Q, and temperature increase. This is evidenced by the decrease in the EISF and the increase of $A_1$. On the other hand, other dynamics, becoming too fast, exceed the instrumental energy window and contribute to the background, as demonstrated by the increase of the background.\\
The width of the Lorentzian, $\Gamma$, characterizes the relaxation time of the dynamics captured within the energy window of the instrument. Interestingly, $\Gamma$ decreases with Q values (Figure S12 in the SI). The temperature dependence is not pronounced. If considered separately, this would mean that there is a confinement effect, i.e. pointing toward faster dynamics at shorter distances. However, given the range of the dynamics of P3HT and the increase in both the Lorentzian and background components with Q value, it is more likely that the distribution of the dynamics is broader than the instrumental energy window. Thus, depending on the Q values, the part of the continuous distribution captured by the spectrometer is not the same. Therefore, we propose that a better approach for these materials is to perform the probe directly in the time-space by measuring the dynamics using a NSE spectrometer. The NSE technique has a two-fold merit, allowing to measure directly $F(q,t)$ and to extend the accessible time scale from picosecond to the nanosecond time domain. This was accomplished using the NSE spectrometer IN11 to probe h-RR-P3HT (Figures \ref{fig:intermediate_200K} and \ref{fig:intermediate_330K}), as well as the backscattering spectrometer IN16B to carry out QENS measurements of d-RR-P3HT (Figures \ref{fig:IN16B_180K} and \ref{fig:IN16B_360K}). IN16B has an elastic resolution of 0.75 $\mu$eV at the used neutron incident wavelength 6.271 \r{A}, which extends the above measured TOF-based time scale from the picosecond to the nanosecond here.
To analyze all the acquired spectra in a consistent way, $F(Q,t)$ was derived from the IN6 spectra of h-RR-P3HT using the following equation:
\begin{equation}
    F(\mathbf{Q},t) = \frac{\int d\mathbf{\omega}  S'(\mathbf{Q},\omega)e^{i\omega t}}{\int d\mathbf{\omega} R(\mathbf{Q},\omega)e^{i\omega t}}
\end{equation}
At shorter time-scales, TOF-based $F(Q,t)$ from IN6, and NSE-based $F(Q,t)$ from IN11, are in a good agreement (Figure S14 in the SI). Note that we are comparing data collected at 250K and 360K on IN6 with measurements carried out at 200K and 330K on IN11, which can account for the small discrepancies, especially at higher Q-values. The NSE-based $F(q,t)$  are also in good agreement with our previous work \cite{Guilbert2015}. At 200K, some dynamics are observed but they are still slow in comparison with the experimental time-scales. At 330K, $F(q,t)$ can be fitted using a simple Kohlrausch-Williams-Watts (KWW) relaxation function with the stretched exponent equal to 0.5 (Table \ref{tab:fitting} and Figure \ref{fig:intermediate_330K}):
\begin{equation}
    F(\mathbf{Q},t) = (1-y_0)exp^{-(t/\tau)^{0.5}}+y_0
\end{equation}
where $\tau$ is the relaxation time. The exponent 0.5 is reasonable for polymeric systems.\cite{Richter2005} This strengthens further the above mentioned statement that these polymers are subject to a continuous distribution of dynamics taking place over a larger time scale.
\begin{table}[H]
\begin{center}
\caption{Parameters and the related standard errors from the fit of $F(q,t)$ from NSE measurements at 330K (Figures \ref{fig:intermediate_330K}), using the KWW model, with a stretched exponent of 0.5.}\label{tab:fitting}
\begin{tabular}{ccc}
\hline
Q (\r{A}$^{-1}$) & $y_0$ & $\tau$ (ps) \\ \hline
0.1 &	0.94 & 	$>$100 \\
0.51 &	0.67  &	70$\pm$26 \\
0.63 &	0.52  &	65$\pm$17 \\
0.85 &	0.42  &	52$\pm$8  \\
0.96 &	0.37  &	31$\pm$4 \\
1.18 &	0.21  &	22$\pm$4 \\ \hline
\end{tabular}
\end{center}
\end{table}
The Q-dependence of the QENS signal for (deuterated) d-RR-P3HT, from the backscattering measurements using IN16B (Figures \ref{fig:IN16B_180K} and \ref{fig:IN16B_360K}) show slower dynamics than the NSE measurements of (hydrogenated) h-RR-P3HT. This is due to the fact that collective motions (incoherent signal is not dominating) are expected to be slower than the self-motions (incoherent signal dominating). The broadening of the peaks increase with temperature, however, although a dependence on Q is observed, it is again less marked than for h-RR-P3HT, especially within the range 1.0-1.4 \r{A}$^{-1}$.\\
The MD simulations allowed us to calculate the  time-evolution of both the distinct and the self-part of the van Hove function as a function of temperature (Figure \ref{fig:Van_Hove}). This is in order to estimate the coherent and incoherent intermediate scattering functions (Figure \ref{fig:Intermediate_scattering_function_MD}) by performing a proper neutron weighting of their contributions, and combining Equations \ref{eq:van_hove_splitting} and \ref{eq:FFT_to_intermediate}. \\
For the crystalline RR-P3HT sample, simulations up to 300K reveal clear peaks in G$_d$(r,t), especially at shorter distances, while two distributions are needed to fit G$_s$(r,t). At 360K, the time dependence is pronounced, the short distance peaks from G$_d$(r,t) vanish, and only one distribution is observed for G$_s$(r,t). Note that this distribution is non-Gaussian, which is in a good agreement with the KWW model used above to describe the relaxation. This change in behaviour is in good agreement with the observed changes in diffractograms for RR-P3HT at higher temperatures.\\
F$_{coh}$(q,t) reflects slower dynamics than F$_{inc}$(q,t) for all the studied temperatures. The self-motion relaxation times decrease with Q and temperatures. Almost no differences between the different models are observed for the self-motions, except at high Q-values and temperatures up to 300K, where the crystalline sample appears to exhibit a slightly slower dynamics. Looking at F$_{coh}$(q,t), a difference in dynamics is observed between the low Q-values and higher Q-values than about 1.4 \r{A}$^{-1}$. For higher Q-values, the dynamics seems to be damped with the effect most visible at lower temperatures for the crystalline RR-P3HT model. At 360K, almost no differences are observed between the different MD simulation models. These changes in dynamics at high Q-values can be linked with the microstructure of the polymers; the dynamics at higher Q-values linked with the $\pi-\pi$ stacking is different than the dynamics at lower Q-values linked with the lamellar stacking. Crystalline models seem more impacted at higher Q-values than amorphous models while no differences between the models can be observed at lower Q-values reinforcing the idea that the side chains are amorphous and that their dynamics is marginally impacted by the backbone conformation and arrangement.
To summarize, we observed no differences in self-relaxation between h-RR-P3HT and h-RRa-P3HT on the picosecond time-scale and no significant differences were calculated on the nanosecond timescale between amorphous and crystalline models. However, collective motions appear different at low and high-Q values with the crystalline models and the d-RR-P3HT most impacted. The differences in motions can be linked with motions taking place in the $\pi-\pi$ direction, and in the lamellar direction.
\paragraph*{Inelastic neutron scattering}
Diffraction, or elastic scattering, probes primarily structural long-range order based on the coherent scattering, and incoherent scattering is present as a background. Quasi-elastic scattering probes a combination of "slow" self-motions and collective motions on picosecond to nanosecond time scale. Vibrational spectroscopy probes "fast" vibrational motions on the femtosecond time scale and provides further information about both long-range order, for instance acoustic modes, as well as short-range order, for instance optical modes, such as segmental orientation and conformational distribution. Neutron vibrational spectroscopy explores both the coherent (phonons or external modes) as well as the incoherent (molecular or internal modes) scattering. Thus, to thoroughly characterize the structural dynamics of our polymers, we explored the lattice/phononic and molecular vibrational sides (discrete excitations, while QENS allows to probe quasielastic continuous energy exchange) by performing inelastic neutron scattering (INS). An INS spectrometer performs an accurate measurement of the dynamical scattering function, S(Q,E), where the evolution of the system is encoded. Vibrationally, the advantage offered by neutrons in comparison with the more widely used Raman and IR spectroscopy, is an appropriate Brillouin zone coverage, the use of deuteration for contrast variation, the absence of selection rules and the absence of photoluminescence from the conjugated polymers. Both the lattice and molecular degrees-of-freedom can be probed using complementary neutron spectrometers - in terms of accessible energy ranges - to cover suitably the associated energy transfer within specific temperature domains.  
\begin{figure}[H]
\begin{subfigure}[b]{0.45\textwidth}
\includegraphics[width=\textwidth]{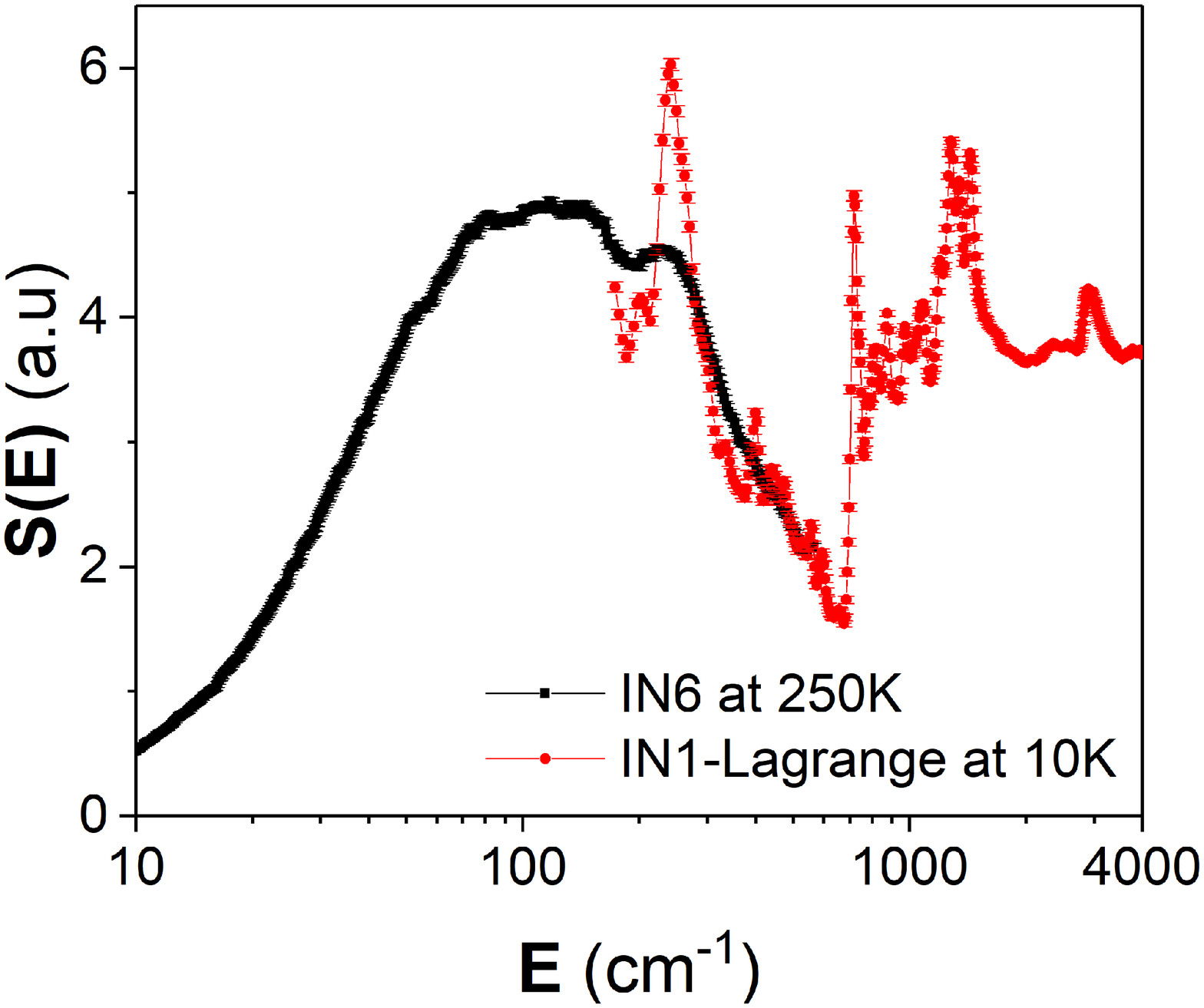}
\caption{}\label{}
\end{subfigure}
\begin{subfigure}[b]{0.45\textwidth}
\includegraphics[width=\textwidth]{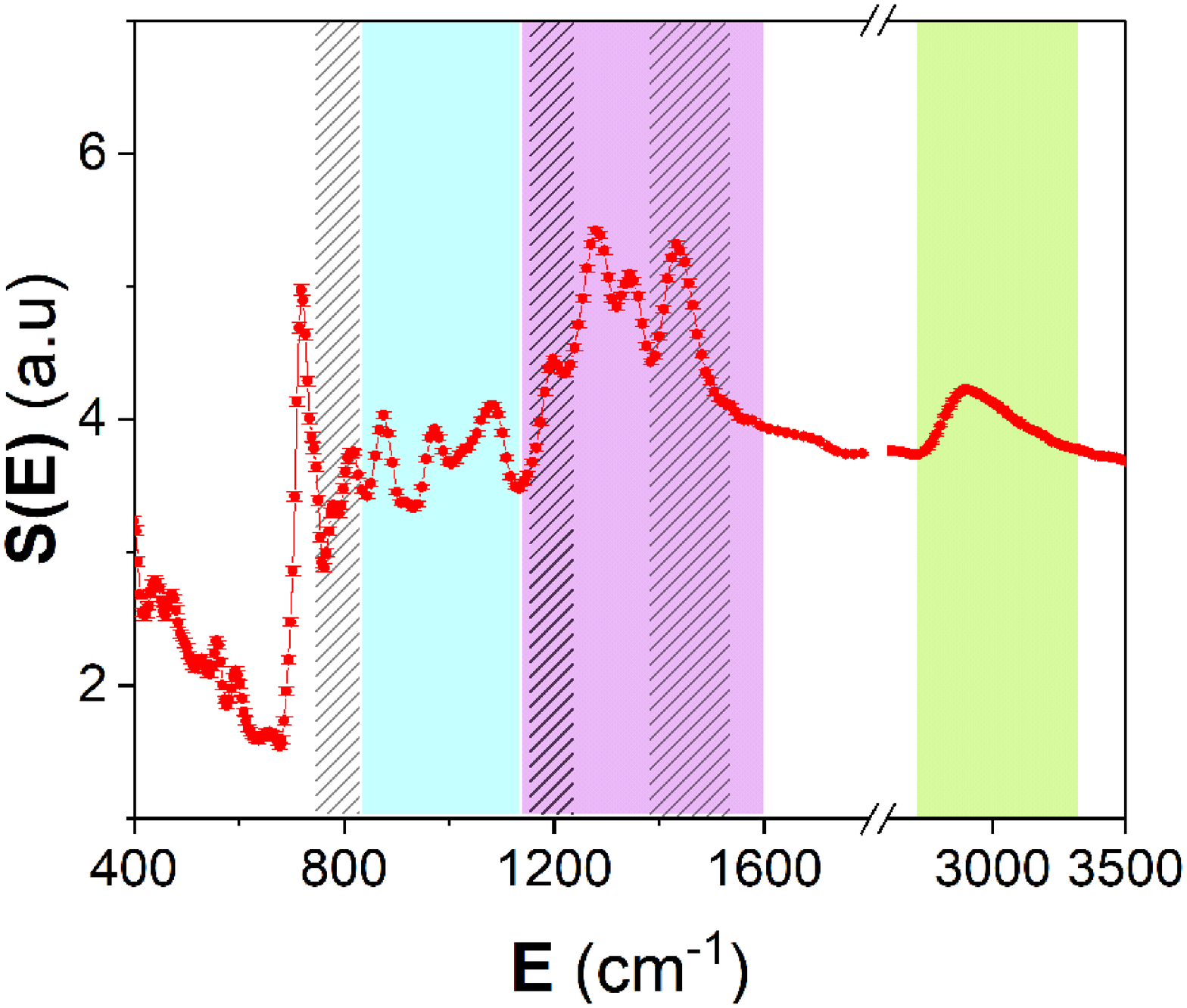}
\caption{}\label{}
\end{subfigure}
\caption{INS spectra of h-RR-P3HT: (a) spectral range overlap and coverage from measurements using IN6, at 250K, and IN1-Lagrange, at 10K, and (b) color shaded illustration highlighting the different molecular mode regimes from measurements on IN1-Lagrange, at 10K.}\label{fig:IN6_vs_IN1}
\end{figure}
Presently, we used the cold-neutron, direct geometry, time-of-flight spectrometer IN6, and the hot-neutron, inverted geometry, IN1-Lagrange. IN6 allows to measure the low-energy modes in the up-scattering regime (anti-Stokes spectra) around room temperature, while IN1-Lagrange is dedicated to probe higher-energy modes covering all the molecular degrees-of-freedom in the down-scattering regime (Stokes spectrum)  at low (base) temperature. In both cases the extracted spectrum is a neutron-weighted \cite{gdosvdos1} energy-distributed density of vibrational (lattice or molecular) modes. Figure \ref{fig:IN6_vs_IN1} illustrates this, and shows the vibrational spectrum of h-RR-P3HT using IN6, at 250K, and IN1-Lagrange, at 10K. The energy transfer overlap between the two spectrometers is highlighted in the range 200-500 cm$^{-1}$. IN1-Lagrange covers the whole high-energy molecular vibrational part of the spectrum. The peaks become less defined and broaden as temperature rises, due to the Debye-Waller effect. For h-RR-P3HT, the C-H stretch band can be observed around 3000 cm$^{-1}$. The energy range 1150-1600 cm$^{-1}$ is associated with the C=C stretching modes, as well as CH$_3$ umbrella motion and CH$_2$/CH$_3$ bending modes in the frequency domain 1350-1500 cm$^{-1}$. The C-H bending mode is observed around 1200 cm$^{-1}$. The range 850-1150 cm$^{-1}$ is associated with mode coupling between the thiophene ring and the side-chains \cite{Brambilla2014}. The peak around 730 and 820 cm$^{-1}$ are associated with the C-S-C  deformation\cite{Tsoi2011}, and C=H out-of-plane deformation modes \cite{Brambilla2014,Brambilla2018}, respectively. The range 400-730 cm$^{-1}$ is associated with C=C out-of-plane deformation modes. The frequency region below 400 cm$^{-1}$ concerns mainly the lattice or external modes. The symmetric C=C stretch mode around 1460 cm$^{-1}$  is predominant in Raman spectra and is often used as a marker of the planarity of the backbone. The medium-weak Raman features centered around 1000 cm$^{-1}$ has been shown to be linked with the conformation of the hexyl side chains. The =C-H out-of-plane deformation mode (820 cm$^{-1}$) is an IR active transition and has been related to both the backbone and side chains conformation \cite{Brambilla2018}. It is worth to note the good match between our INS spectra and the available Raman/IR data in literature.\cite{Brambilla2014,Tsoi2011,Brambilla2018}
\begin{figure}[H]
\begin{subfigure}[b]{0.45\textwidth}
\includegraphics[width=\textwidth]{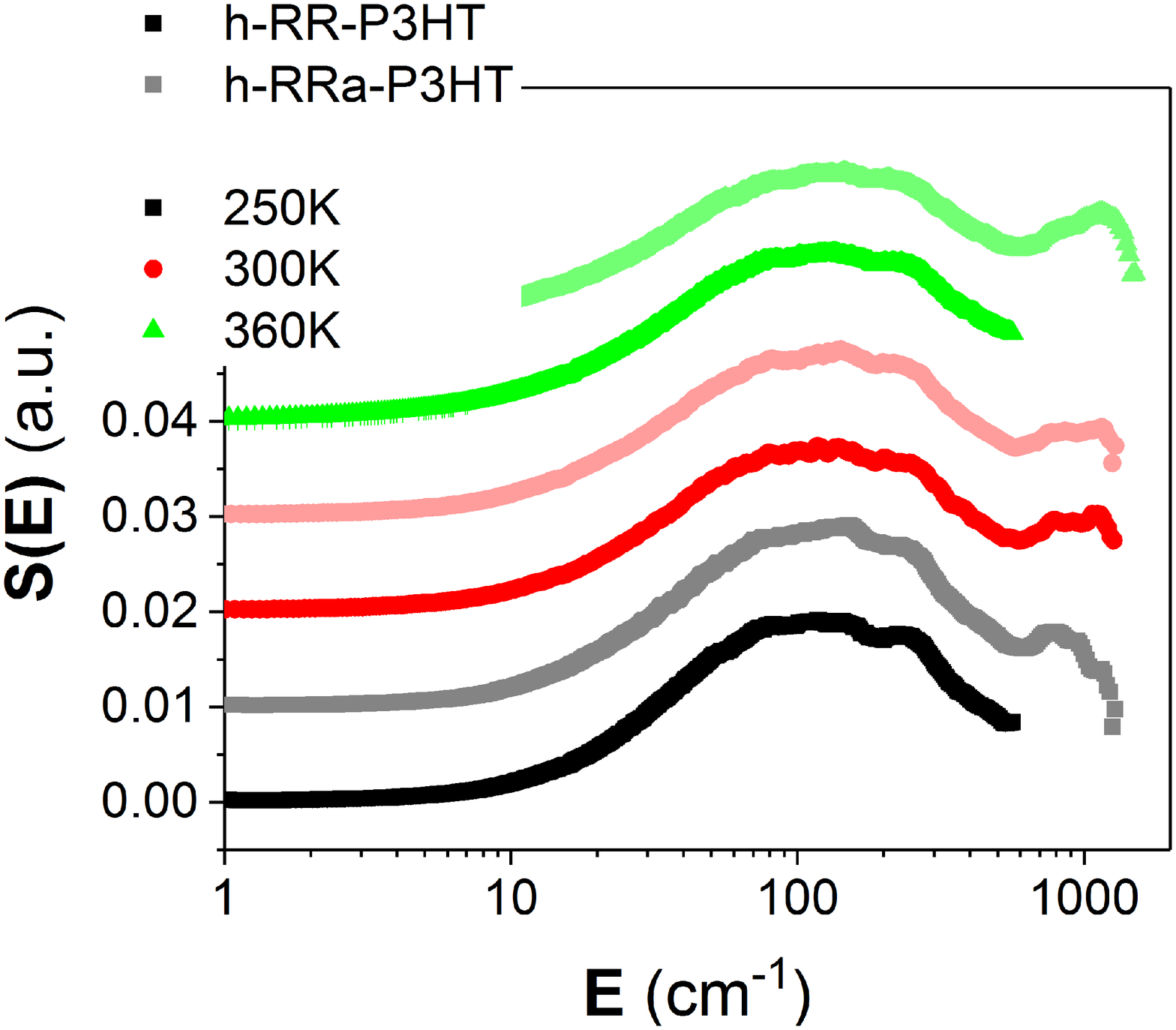}
\caption{}\label{fig:IN6-h-DOS}
\end{subfigure}
\begin{subfigure}[b]{0.45\textwidth}
\includegraphics[width=\textwidth]{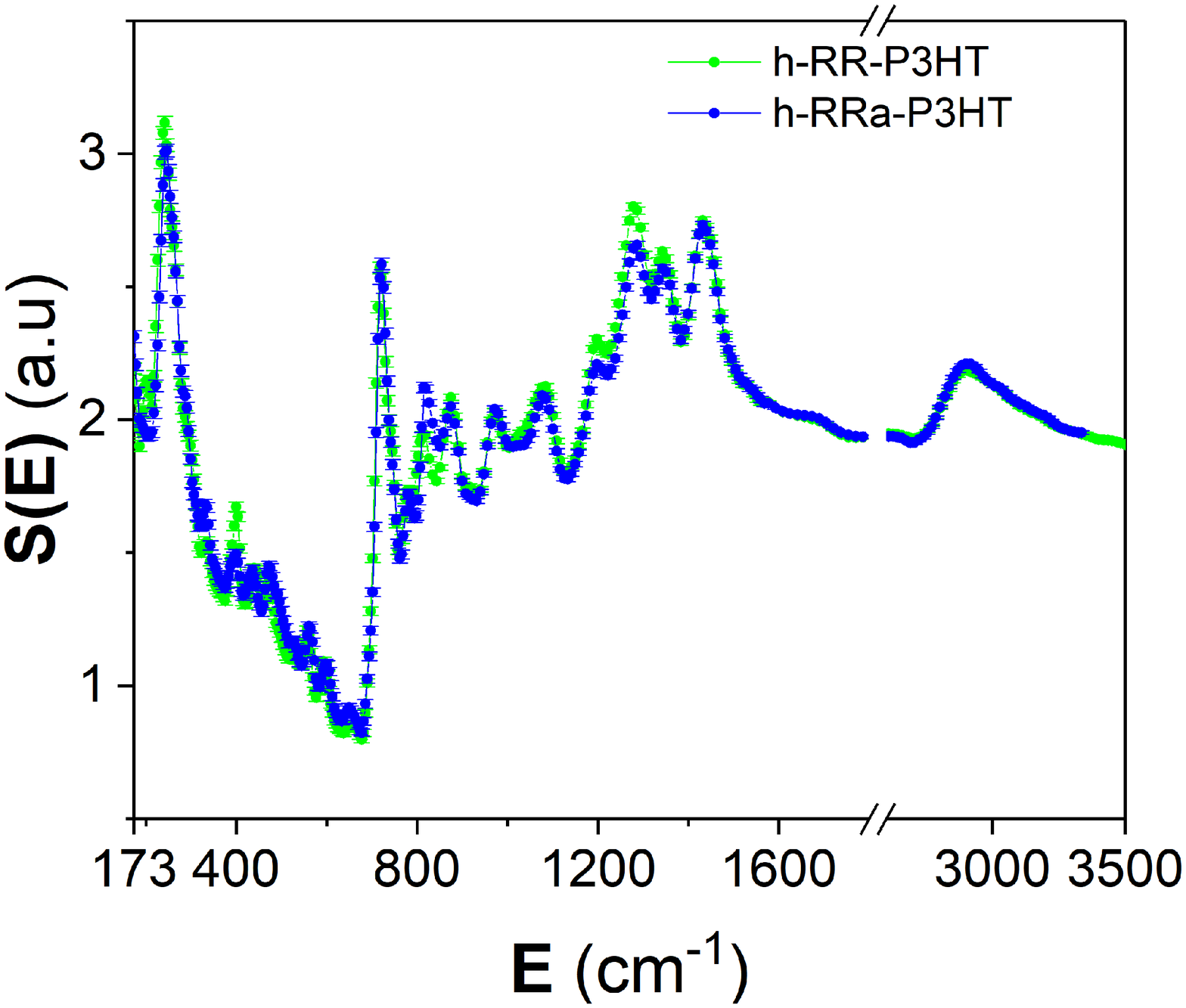}
\caption{}\label{fig:IN1-h-DOS}
\end{subfigure}
\caption{INS spectra of h-RR-P3HT and h-RRa-P3HT measured on (a) IN6 at different temperatures around T$_g$, and on (b) IN1-Lagrange at 10K.}\label{fig:h-INS}
\end{figure}
Figure \ref{fig:h-INS} presents the INS spectra of h-RR-P3HT and h-RRa-P3HT from measurements using both the cold-neutron IN6 and hot-neutron IN1-Lagrange spectrometers. Only subtle differences can be observed between both materials when comparing their vibrational spectra.\\
On IN6 (Figure \ref{fig:IN6-h-DOS}), data were collected around T$_g$, at 250, 300 and 360K, at the anti-Stokes side, and covers the low-energy part, mostly related to lattice degrees-of-freedom (external modes). Upon heating from 250 to 360K, across the T$_g$, the spectra become less defined. For instance the phonon band around $\sim$ 250 cm$^{-1}$ broadens and mergers with the band at $\sim$ 150 cm$^{-1}$ as temperature rises. Spectra are broad given the soft nature of the systems, the large volume of amorphous material and the broad numbers of conformations that can be explored by the materials at those temperature. The lattice component of both the samples is found to exhibit a closely similar vibrational behavior. This leads to assume that the considered amount of regioregularity has no pronounced effect on the description of the external modes of hydrogenated P3HT in the amorphous phase dominating the INS signal.\\
The main interest resides in the molecular description of both RR-P3HT and RRa-P3HT. IN1-Lagrange (Figure \ref{fig:IN1-h-DOS}) allows to map out fully all the internal (molecular) modes of both the materials. The peak at $\sim$ 250 cm$^{-1}$ and 800 cm$^{-1}$ are slightly shifted towards low-energy for h-RR-P3HT. The intensity of the peaks at 330 cm$^{-1}$ and 800 cm$^{-1}$ is larger for h-RRa-P3HT while the intensity of the peaks at 400, 1200, 1280, and 1350 cm$^{-1}$ is higher for h-RR-P3HT. The peak around 1080 cm$^{-1}$ is more intense for h-RR-P3HT and a shoulder at lower energy (~1030 cm$^{-1}$) is present for h-RR-P3HT. The changes in the intensity of the peaks in the range 1200-1400 cm$^{-1}$ and in the feature around 1080 cm$^{-1}$ and 800 cm$^{-1}$ can be interpreted as slightly more planar backbones and a slight ordering of the hexyl side chains for h-RR-P3HT, predominantly in the crystalline phase. The changes observed below 700 cm$^{-1}$ are thus likely linked with h-RR-P3HT being slightly stiffer than h-RRa-P3HT. Note that neutron spectroscopy is very sensitive to hydrogen and thus, the Raman active symmetric C=C stretch mode is likely to be covered here by the CH2/CH3 bending mode. 
\begin{figure}[H]
\begin{subfigure}[b]{0.45\textwidth}
\includegraphics[width=\textwidth]{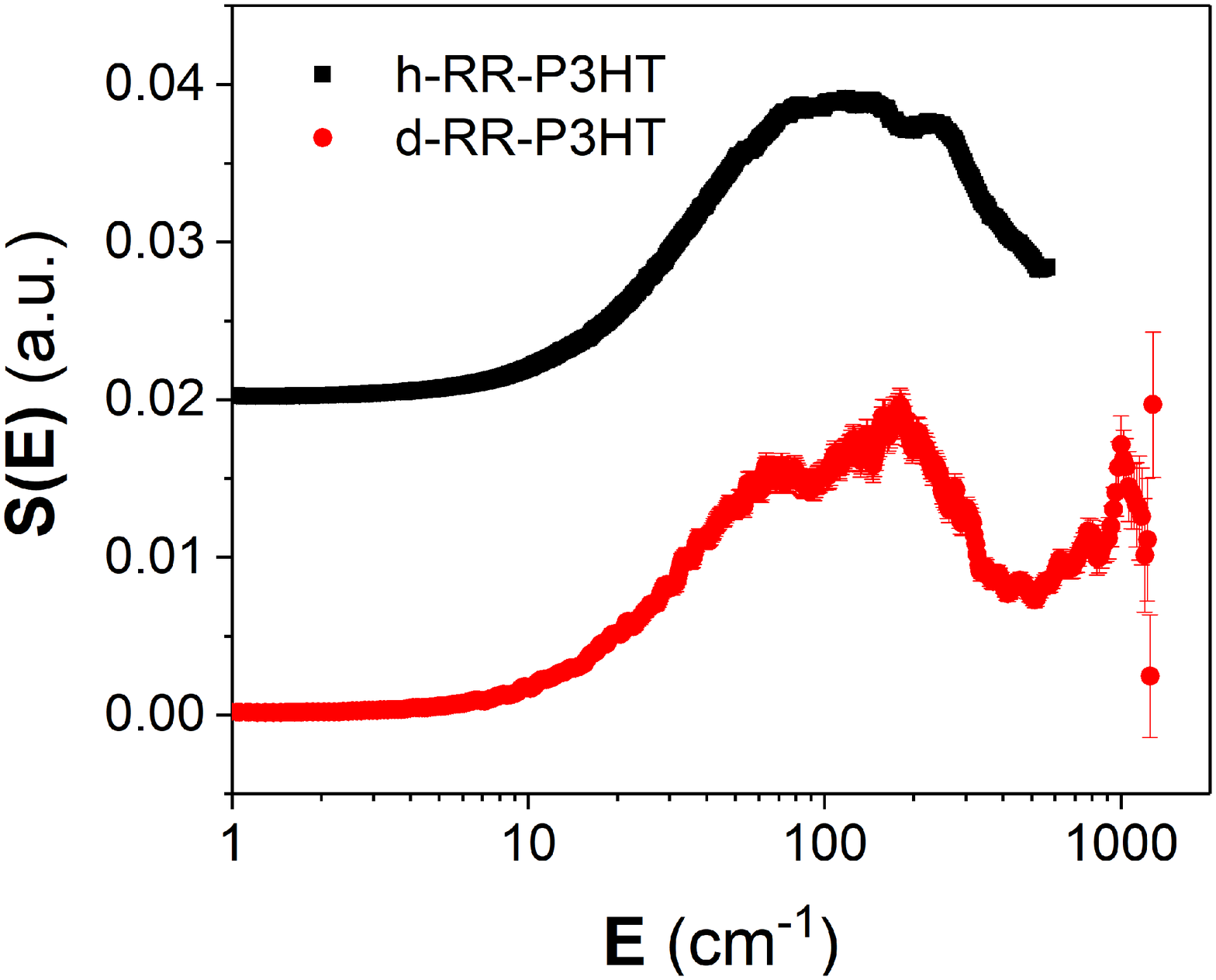}
\caption{}\label{fig:IN6_h_RR_vs_d_RR}
\end{subfigure}
\begin{subfigure}[b]{0.45\textwidth}
\includegraphics[width=\textwidth]{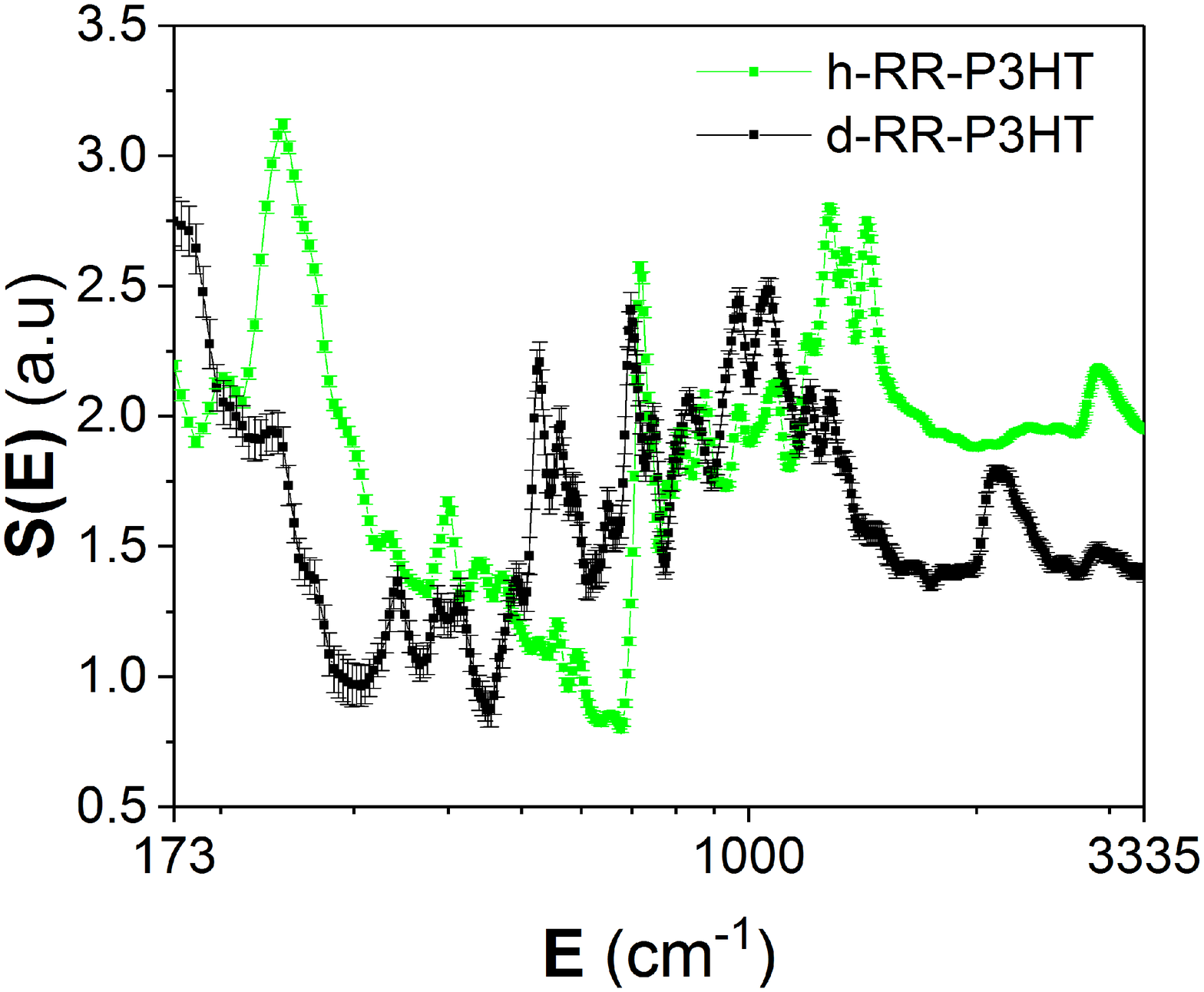}
\caption{}\label{fig:IN1_h_RR_vs_d_RR}
\end{subfigure}
\begin{subfigure}[b]{0.45\textwidth}
\includegraphics[width=\textwidth]{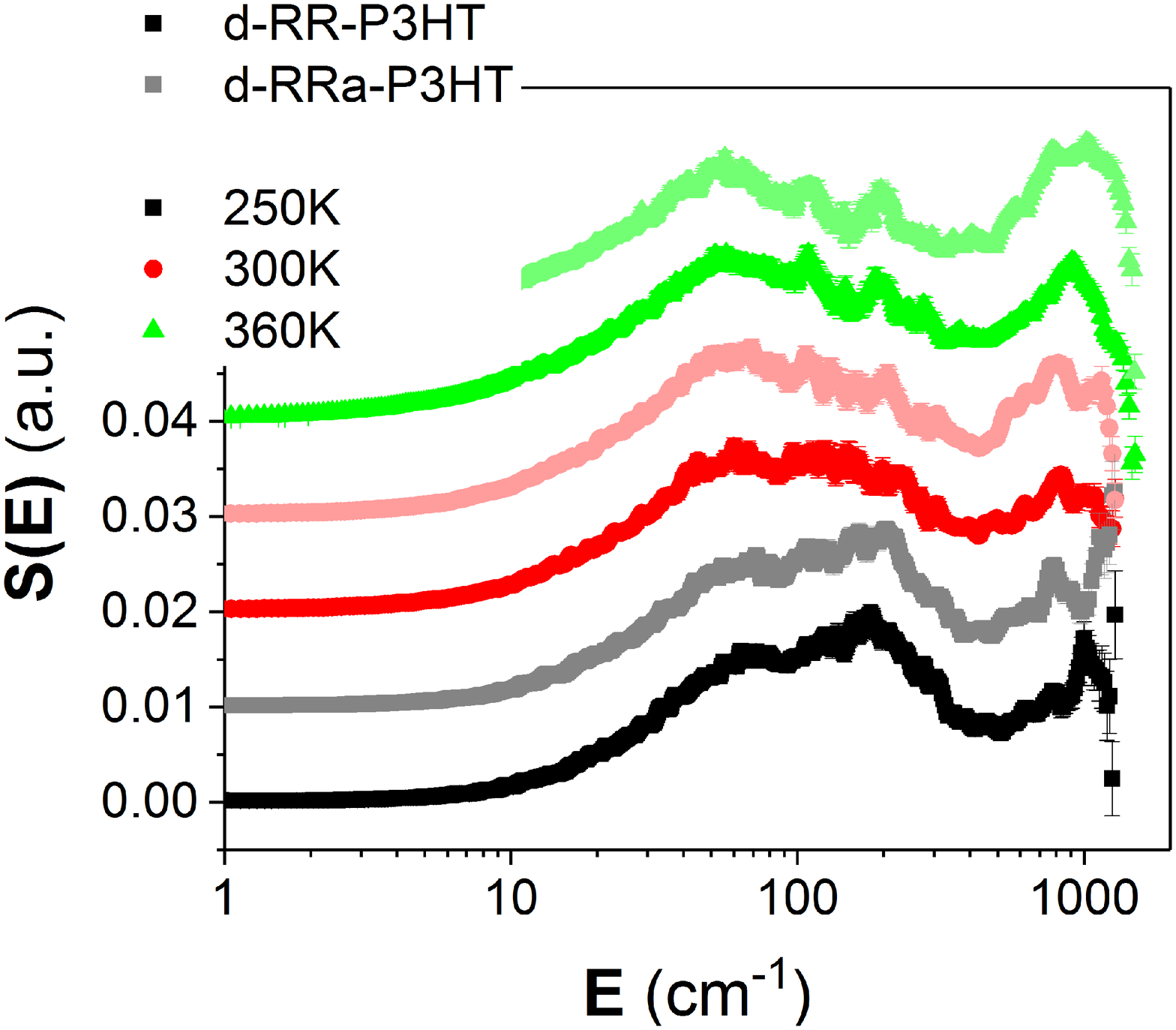}
\caption{}\label{fig:IN6-d-DOS}
\end{subfigure}
\begin{subfigure}[b]{0.45\textwidth}
\includegraphics[width=\textwidth]{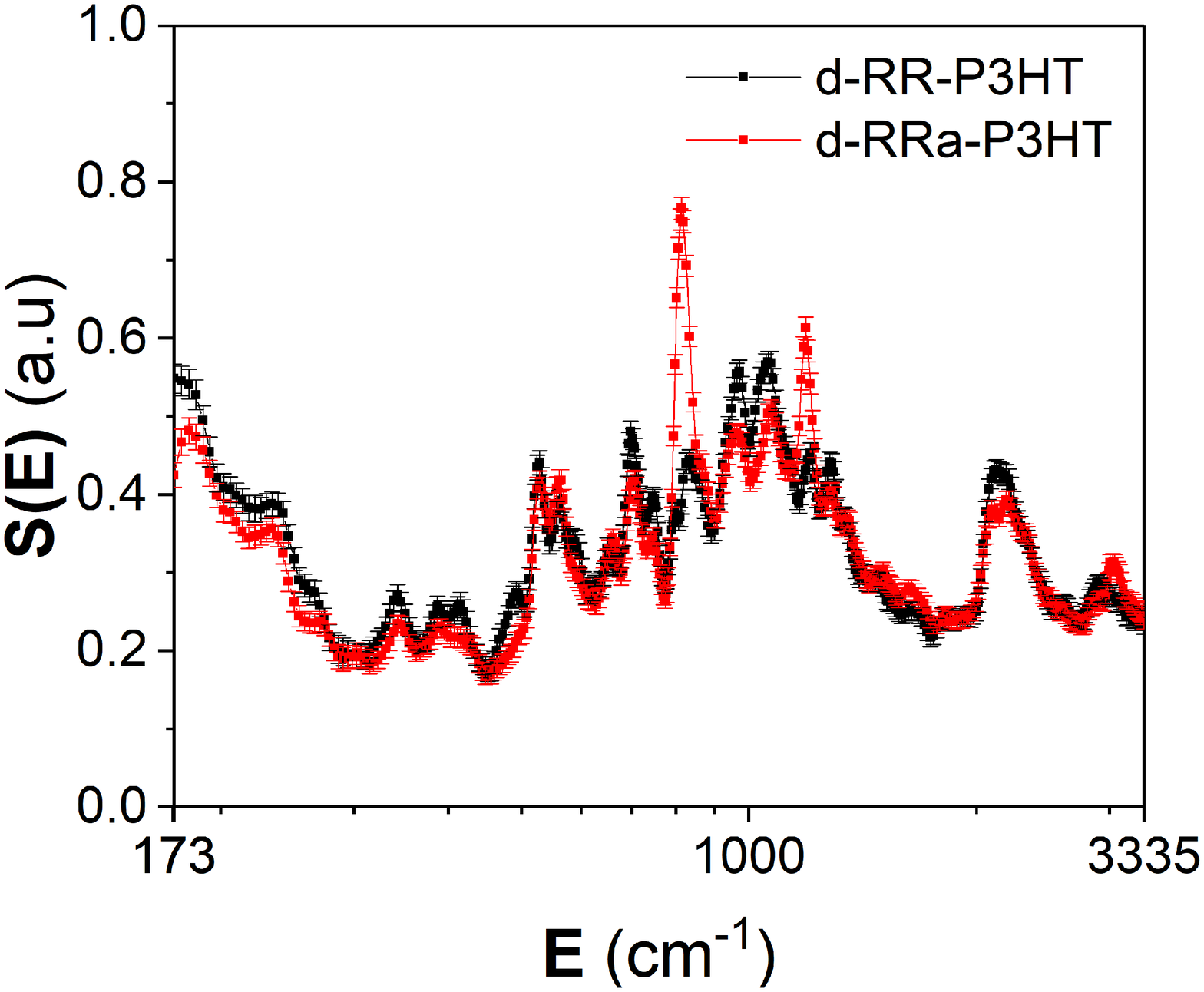}
\caption{}\label{fig:IN1_d_RR_vs_d_RRa}
\end{subfigure}
\caption{Top panel: INS spectra of h-RR-P3HT and d-RR-P3HT from measurements on (a) IN6, at 250K, and (b) IN1-Lagrange, at 10K. Bottom panel: INS spectra of d-RR-P3HT and d-RRa-P3HT measured on (c) IN6 at different temperatures, around T$_g$ and (d) IN1-Lagrange, at 10K. }\label{fig:d-INS}
\end{figure}
Deuteration allows to better highlight vibrational features not dominated by, or, not involving dynamics of hydrogen atoms. Figure \ref{fig:IN6_h_RR_vs_d_RR} compares measured GDOS of h-RR-P3HT and d-RR-P3HT, at 250K, using IN6. The spectrum of the deuterated sample (d-RR-P3HT) is better defined, highlighting features around 70, 200 and 1000 cm$^{-1}$, which were not clearly observed in the hydrogenated sample (h-RR-P3HT). Considering this contrast variation offered by deuteration, the effect of regioregularity can be investigated. The temperature evolution of GDOS of d-RR-P3HT and d-RRa-P3HT, using IN6, is presented in Figure \ref{fig:IN6-d-DOS}. Similarly to the GDOS of the hydrogenated samples, h-RR-P3HT and h-RRa-P3HT, discussed above (Figure \ref{fig:IN6-h-DOS}), vibrational spectra of d-RR-P3HT and d-RRa-P3HT do not exhibit noticeable differences that can be associated with the difference in regioregularity. However, a temperature dependence is observed upon heating from 250 to 360K, across the T$_g$. The phonon band around 70 cm$^{-1}$ sharpens, and its intensity increases in comparison with the feature around 200 cm$^{-1}$. The sharp vibrational band at $\sim$ 1000 cm$^{-1}$ becomes more pronounced and broadens as temperature rises, due to population increase and Debye-Waller effect, respectively.\\
Deeper insights are gained by mapping out the full energy range of the internal modes, covering the vibrational molecular degrees-of-freedom. Deuterated samples were also probed using IN1-Lagrange. Upon deuteration, a mass effect can clearly be observed (Figure \ref{fig:IN1_h_RR_vs_d_RR}). This is due to the mass of the deuterium being the double of that of the hydrogen. In this context it can be seen that most peaks of d-RR-P3HT are red-shifted when comparing to h-RR-P3HT (Figure \ref{fig:IN1_h_RR_vs_d_RR}). This effect is not linear.i.e it is much pronounced at higher energies where the C-H and C-D stretch modes are contributing the most. Note that modes not involving directly H/D will only be slightly impacted. A shift to lower energies of $\sim$ 800 cm$^{-1}$ is observed when going from fully hydrogenated to almost fully deuterated samples. In fact, a reminiscent of C-H stretch is observed for both deuterated samples (Figure \ref{fig:IN1_d_RR_vs_d_RRa}). This is due to end-group of polymers as well as defect in the exchange of hydrogen with deuterium on the backbone (Figure S1-S7 in Supporting Information). The large difference in intensity for the peaks at 800 cm$^{-1}$ and 1200 cm$^{-1}$ is due to the hydrogen defects present on the backbones of h-RRa-P3HT. This confirms that these peaks can be assigned to the =C-H out-of-plane deformation mode and the C-H bending mode, respectively. It also highlights the potential of using partially-deuterated materials. Although the intensities of all the peaks are modulated by the presence of these hydrogen defects, the most noticeable changes are the presence of additional peaks at 490 cm$^{-1}$ and 590 cm$^{-1}$ for d-RR-P3HT. A further finding worth of noting, as made to be easily observed by deuteration, is the peak splitting seen in some of the peaks in deuterated samples. This splitting was of a weak nature or masked in the hydrogenated samples. It concerns mainly vibrational bands below 1000 cm$^{-1}$, and mostly due to specific motions not involving hydrogen atoms, like the C-S-C deformation ($\sim$ 730 cm$^{-1}$), and could also be due to inter-molecular interactions induced by or coupled to the lattice degrees-of-freedom. \\
Note that the predominant Raman-active peak is hardly visible in the INS spectra for the deuterated materials, and if it would be present, it would largely be masked by the CH$_2$/CH$_3$ bending mode in the hydrogenated samples. This observation points towards the idea that this predominant Raman-active peak is linked with collective skeletal stretching, and is strongly affected by $\pi$-electrons delocalization\cite{Brambilla2014}. Thus, highlighting the appropriate complementarity of INS and optical spectroscopy techniques such as Raman.\\
To summarize, the effect of regio-regularity is not pronounced experimentally, when comparing RR-P3HT and RRa-P3HT. From both IN6 and IN1-Lagrange measurements (Figure \ref{fig:IN1_c}), the spectra of the h-RR-P3HT and h-RRa-P3HT resemble each other, but subtle differences are captured, although being of a weak nature. This points towards equivalent intra-molecular modes, and a closely similar lattice dynamical behavior. This is also the case when the strong incoherent scattering from hydrogen atoms is strongly reduced via deuteration (Figure \ref{fig:IN1_d}), where some differences are better marked. The INS measurements indicate a dominant averaged dynamics of the atomic arrangements of the amorphous phase in both RR-P3HT and RRa-P3HT.\\
\begin{figure}[H]
\begin{subfigure}[b]{0.45\textwidth}
\includegraphics[width=\textwidth]{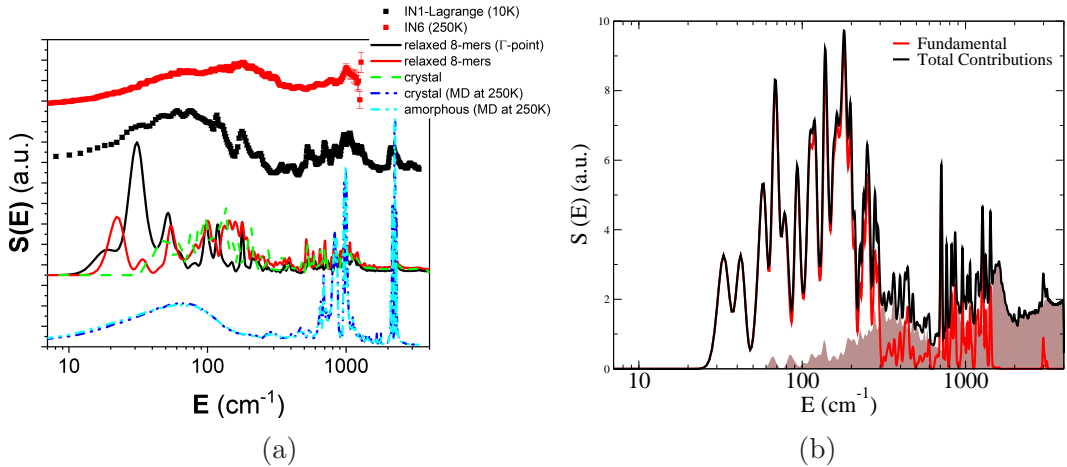}
\caption{}\label{fig:different_calculs}
\end{subfigure}
\begin{subfigure}[b]{0.40\textwidth}
\includegraphics[width=\textwidth]{RR-8_conc.eps}
\caption{}\label{fig:contributions}
\end{subfigure}
\caption{(a) Comparison between measured and calculated INS spectra of d-RR-P3HT. The measured spectra are from IN6 at 250K, and IN1-Lagrange at 10K. The calculated spectra are obtained adopting a combined computational approach, using different numerical techniques, applied to a 8-monomer-based model system (8-mers). These approaches are single molecule DFT ($\Gamma$-point) and periodic DFT.  Crystal periodic DFT was also carried out (crystal). Further classical MD was performed at 250K, on both crystal and amorphous models, where boxes of 100 chains of 20 monomers (20-mers) were considered. (b) Calculated INS spectrum of the hydrogenated 8-mers from periodic DFT calculations. The calculated total spectrum (black) is presented as the sum of the (main) fundamental vibrational transition (red), and other contributions (brown shaded area). The other contributions include higher order transitions and multi-phonons.}
\end{figure}
In order to get a deeper insight into the vibrational signature of the polymers, we adopted a combined computational approach for the sake of complementarity, to better describe the complex structural and dynamical nature of RR-P3HT and RRa-P3HT. Computationally, the aim is to unravel the impact of the polymer conformation, the crystallinity and the impact of the environment on the calculations to better describe the measurements (Figure \ref{fig:different_calculs}).\\
The computational techniques include density functional theory (DFT) and classical molecular dynamics (MD) calculations. Both the single or isolated molecule and periodic calculations were carried out within the framework of DFT. The single or isolated molecule approximation allowed to focus mainly on the intra-molecular vibrational degrees-of-freedom. On the other hand, DFT-based periodic calculations include the lattice dynamical component (external modes), and eventually, its possible interplay and/or coupling with the molecular vibrations (internal modes). Both the approaches allow also to investigate the strength of intra-molecular with respect to inter-molecular vibrational degrees-of-freedom, which is important, for instance, when dealing with correlated peak splitting. Classical MD simulations are also relevant to the present study as MD allows to perform finite-temperature calculations, on larger models, to follow properly both the time and thermodynamical evolution of the systems, offering the advantage to map out appropriately the different structural configurations and conformations, whose dynamics possess similar energies. This averaged description is important in order to reproduce specific characteristics, like the experimental occurrence of a broadening of some features. \\
From the classical MD simulation we extracted the velocity autocorrelation function (VACF) to derive vibrational spectra. Whereas, From DFT-based isolated-molecule and periodic simulations, these were obtained from normal modes and phonon calculations, respectively. In order to compare directly with the measured INS spectra, the calculated partial vibrational contributions of the different atoms were neutron weighted \cite{gdosvdos1}. In addition to estimating the fundamental vibrational transition, higher harmonics as well as phonon wings were also calculated (Figure \ref{fig:contributions}), to better reproduce and describe the measured INS spectra using IN1-Lagrange.\\
Figure \ref{fig:different_calculs} shows that the DFT-based molecular, $\Gamma$-point, approach reproduces reasonably well the intra-molecular modes (high energy regime). It is found to be a quick and efficient method to qualitatively describe the molecular internal modes (Figure \ref{fig:IN1}). However, this isolated molecule approximation does not include the lattice or external degrees-of-freedom, thus not accounting for a possible coupling of the internal molecular modes with external lattice modes, or any eventual short distance, inter-molecular, interaction, i.e. modes coupling or interplay between molecules within the same unit cell. The latter, termed as correlation field splitting, can in principle be studied as a marker for crystallinity of a sample. DFT-based periodic approach allows to correct some of those issues and the main improvement while comparing with INS spectra (Figure \ref{fig:IN1}) occurs in the mid-energy range when the molecular modes are couple with nearby molecules. However, this approach still suffer from the fact that the calculation is done on one possible conformation of the molecule and that this conformation is repeated using periodic boundary conditions. Finally, MD simulations allow to probe the evolution of various conformations in time in the solid-state, leading to an improvement of the low-energy range where the broad peak around 100 cm$^{-1}$ is reproduced (Figure \ref{fig:IN1}). However, MD does not describe well the high-energy range largely because it relies on a classical description (simple oscillators) of all the degrees-of-freedom of the molecules. Note that the MD crystalline model does not reproduce the small features observed at low energy but the periodic DFT approach gives an insight into these features. Thus, by using a combination of these methods, we can describe the entire energy range of the spectra.\\
Indeed, basically all our DFT computational approaches reproduce well the intra-molecular high-energy region (from $\sim$ 800 cm$^{-1}$ and from $\sim$ 500 cm$^{-1}$ for for the protonated and deuterated samples, respectively) of the spectra of both protonated and deuterated polymers (Figure \ref{fig:IN1_a} and Figure \ref{fig:IN1_b}). The lower energy-range is more complex due to the occurrence of inter-molecular coupling and impact of the environment and some features such as the low-energy peak ($<$ 170 cm$^{-1}$) is best reproduced by MD simulation with a clear shift to lower energies and a narrowing of the peak with deuteration (Figure \ref{fig:IN1_a} and Figure \ref{fig:IN1_b}). The peak between 170 cm$^{-1}$ to $\sim$ 500 cm$^{-1}$, especially the peak at 250 cm$^{-1}$ for the protonated samples are relatively well reproduced by the periodic DFT approach, although the smaller peaks in the region between $\sim$ 400 cm$^{-1}$ to $\sim$ 500 cm$^{-1}$ are also captured by MD simulations, highlighting the importance of averaging over a large number of conformations.\\
The model calculations help to shed some light on the averaged picture behind the subtle differences observed from the INS measurements between RR-P3HT and RRa-P3HT. For the protonated calculations, the high-energy modes are more defined for h-RR-P3HT in both DFT approaches. The intensity of the C-S-C deformation mode at $\sim$ 730 cm$^{-1}$ with respect to the intensity of higher-energies mode is better captured by the periodic approach though. The calculated peak at $\sim$ 820 cm$^{-1}$ in the case of h-RR-P3HT is shifted to 800 cm$^{-1}$ when calculating h-RRa-P3HT, and is assigned to modes predominantly linked with the backbones. The molecular approach seems to reproduce better the broadening of the peaks at energies higher than 900 cm$^{-1}$, especially the more intense features at $\sim$ 1000 cm$^{-1}$ in h-RR-P3HT is reproduced. We assign the latter to mode coupling between the thiophene ring modes and the side chains modes. At lower energies, the features below 150 cm$^{-1}$ and at 235 cm$^{-1}$ are well reproduced by the MD calculations and the periodic DFT approach, respectively. The 150 cm$^{-1}$ is not affected by regio-regularity nor by the choice of the models (amorphous or crystalline) in line with the INS measurements. Although the energy range above 170 cm$^{-1}$ is better described by the periodic DFT approach, the features at 450-500 cm$^{-1}$ calculated by MD are shifted towards lower energies for the crystalline model, and similar changes are observed in the INS measurements pointing towards the impact of crystallinity in this energy region. Concerning the high-energy modes in the deuterated calculations, the main differences between RR-P3HT and RRa-P3HT can be observed around 550, 650 and 800 cm$^{-1}$, where the peaks are more intense for d-RR-P3HT. The first two peaks are assigned to coupling between thiophene modes and side chains modes. The last peak is calculated at 820 cm$^{-1}$ for h-RR-P3HT is shifted to 800 cm$^{-1}$ for calculations on h-RRa-P3HT, and is assigned to modes predominantly linked with the backbones. Note we see the same in the protonated samples. The coupling between modes involving both thiophenes and side chains are calculated for the deuterated samples at lower energies ($\sim$ 400 cm$^{-1}$) than the protonated samples as these modes involved directly the hydrogens/deuteriums and are therefore directly impacted by the mass effect, although to a lesser extend than the C-H stretch. At lower energies, the MD calculations reproduce well the broad peak below 170 cm$^{-1}$, and the peak around 170 cm$^{-1}$ is well reproduced by the DFT with the periodic approach. The main differences between d-RR-P3HT and d-RRa-P3HT in the low-energy range occurs at 250, 300, 350 and 390 cm$^{-1}$ where the peaks are broader for d-RRa-P3HT in the periodic DFT calculations. These peaks are also present in the MD calculations and are more intense for the crystalline model, which is in good agreement with the INS measurements.\\
To summarise, the effect of regio-regularity on the vibrational spectra is not pronounced experimentally. Deuteration highlights better some subtle differences. The MD simulations show no noticeable effect of regioregularity, in line with the INS measurements, hence providing a further support of the closely similar averaged conformational motions of both RR and RRa P3HT. Focusing on specific conformational models helps to better spot differences, especially these are well pronounced in the low-energy region, where the lattice and inter-molecular degrees-of-freedom are dominating. The lattice component is not present in the isolated molecule approach, and allows to see that periodic calculations, including the lattice part, lead to a better monitoring of the different phases of the polymer. However, subtle changes in conformation are assigned to changes in intra-molecular modes at higher energies i.e. peaks around 800 cm$^{1}$ assigned to modes linked with the backbones as well as peaks around 1000 cm$^{1}$ for the protonated samples and around 600 cm$^{1}$ for the deuterated samples that are assigned to coupling between thiophene and side chains modes. Therefore, the use of these different computational approaches, in a complementary way, are essential to understand the complex phase-averaged dynamical behavior of the studied conjugated polymer.\\
\section{Conclusions}
We have fully mapped out the structural dynamics of RR-P3HT on a length scale up to 10s of \r{A}, and on the femtosecond to nanosecond time scale. We used a combination of elastic and spectroscopic neutron scattering techniques, and we went a step further with respect to previously reported works in the literature, which were focused on using neutron spectroscopy to predominantly study incoherent self-relaxations and vibrations. Presently, by using deuteration, we gained access to the collective motions and coherent lattice vibrations of the polymers. Thus, we were able to link directly structural features probed by neutron diffraction, mainly the $\pi-\pi$ stacking and lamellar stacking, with different dynamics. \\
Quasi-elastic neutron scattering (QENS) evidenced different slow collective dynamics on an extended picosecond-nanosecond timescale, associated to the $\pi-\pi$ stacking and lamellar stacking. On the picosecond timescale, no differences in terms of self-motions were observed between RR-P3HT and RRa-P3HT. On the other hand, small differences in terms of collective motions were observed, especially for the low Q-values where diffraction features were observed only for RR-P3HT.\\ 
Inelastic neutron scattering (INS) allowed to bridge the gap between molecular conformations and microstructure. Only small differences are observed between h-RR-P3HT and h-RRa-P3HT, and we ascribed them to a slight increase in backbone planarity, and hexyl side chains ordering in the crystalline phase of h-RR-P3HT. The large differences observed between d-RR-P3HT and d-RRa-P3HT were assigned to hydrogen defects present on the backbones, hence highlighting the robustness of partial deuteration. Comparing INS with available Raman and IR spectra from the literature,\cite{Brambilla2014,Tsoi2011,Brambilla2018} it can be emphasized that the intense Raman peak around 1460 cm$^{-1}$ (C=C stretch) is linked with $\pi$-electrons delocalization, as it is not observed in the INS spectra of deuterated P3HT. This strengthens further the adequacy of the use of the INS technique as to also complement efficiently optical spectroscopy techniques like Raman and IR, especially when dealing with semiconducting materials.\\
In order to disentangle coherent and incoherent contributions to the measured neutron scattering spectra, as well as to study the various phases of RR-P3HT, we made various crystalline and amorphous models of RR-P3HT and RRa-P3HT, and used a combination of classical MD simulations and DFT-based quantum chemical calculations. The diffraction patterns as well as the slow relaxations calculated from the relaxed crystalline and amorphous MD models were in a good agreement with the measurements. The simulations point towards small differences on a length scale related to the $\pi-\pi$ stacking, and on the picosecond-nanosecond time scale between amorphous RR-P3HT and RRa-P3HT, as RR-P3HT chains seems to exhibit a preferred orientation, and the dihedral distribution between the monomers is different, resulting in a slightly larger radius of gyration for RRa-P3HT. These small changes of conformations are captured by the INS measurements and can to some extent be captured by quantum chemical calculations. Molecular quantum chemical (QC) calculations give accurate reproduction of the internal modes at high energies, where we found that stretching and bending modes are dominating. In the mid-energy range, periodic QC calculations produced a broadening of the peaks obtained using the isolated molecule QC approximation. These vibrational bands are predominantly associated with coupled modes between side chains and monomers, as well as between adjacent monomers along the backbone (intra-chain) or between nearby chains (inter-chain). The lowest energy part of the spectra is best reproduced by MD simulations, although no noticeable differences are observed between the different models, and the features linked with crystallinity could not be reproduced. The combined neutron-based experimental and multi-computational approaches were found to complement synergistically each other, by covering appropriately different length scales and time scales, and by describing adequately the different phases and energy landscapes of RR-P3HT and RR-P3HT.\\
We conclude that microscopically - via diffraction measurements and by probing slow relaxations - RRa-P3HT is a good approximation for the amorphous phase of RR-P3HT, although at the molecular level - via molecular vibrations and vibrational coupling - some differences between regio-regular and regio-random P3HT are observed even in the amorphous phase.

\section{Experimental}

\subsection*{Sample preparation}
h-RR-P3HT was obtained from Merck Chemicals and h-RRa-P3HT was obtained from Sigma-Aldrich. d-RR-P3HT was synthesized as previously described,\cite{Guilbert2016} while d-RRa-P3HT was synthesized by an iron(III) chloride mediated oxidative polymerisation of 4-d$_1$-3-d$_{13}$-hexylthiophene in chloroform at room temperature as further detailed in the Supporting Information.\\
The molecular weights, polydispersities and regioregularities are summarised in Table \ref{tab:SLD-2}. Regioregularities of deuterated polymers were estimated from $^1$H-NMR analysis of hydrogenated polymers synthesized under identical conditions as further detailed in the Supporting Information.\\ 
The as-received materials were dissolved in chloroform (40mg/mL) and drop-cast on a glass slide on a hot plate at 60°C for an hour. The drop-cast films were then scratched from the glass substrates and stacked in aluminum. Each measured samples were about 400 mg.
\subsection*{Neutron Diffraction}
Neutron diffraction on the scratched drop-cast films was performed using the D16 diffractometer at the Institut Laue Langevin (ILL) in France. A neutron incident wavelength of 4.52 \AA \ with a wavelength spread  of  $\Delta\lambda/\lambda$ = 0.01 was selected to ensure a good compromise between d-spacing range and angular resolution. The beam was collimated by a pair of slits fitting the beam size on the sample and reducing the background. The sample was loaded in a sealed Vanadium cell. The diffracted beam was measured over an angular range of 2-120$^{\circ}$ to have a q range of 0.04 – 2.4 \AA$^{-1}$. The neutron data were corrected for the empty cell, the ambient room background and the non-uniform detector response, and the transmission and the thickness of the sample were also taken into account. The scattering intensity were normalized in absolute unit with a standard calibration.
\subsection*{Quasielastic Neutron Scattering}
The QENS measurements on hydrogenated and deuterated
RR and RRa P3HT (prepared as described above) were performed on the direct geometry cold neutron, time-of-flight, time-focusing spectrometer IN6, and on the backscattering spectrometer IN16B, at the Institut Laue Langevin (Grenoble, France). A small sample thickness of 0.2 mm was used, as it is relevant to the minimization of effects like multiple scattering and absorption.\\ 
On IN6, data were collected at 250, 300 and 360 K, in such way to cover suitably the glass transition temperature region of the polymer. An incident neutron wavelength $\lambda_i$ = 5.12 \AA (E$_i$ = 3.12 meV) was used, offering an energy resolution at the elastic line of $\sim$ 0.07 meV. Standard corrections including detector efficiency calibration and background subtraction were performed. A standard vanadium sample was used to calibrate the detectors and for an instrumental resolution purpose. The data analysis was done using ILL software tools. At the used wavelength, the IN6 angular detector coverage ($\sim$ 10-114\degree) corresponds to a Q-range of $\sim$ 0.2-2.1 \AA$^{-1}$. Different data sets were extracted either by performing a full Q-average in the (Q,E) space to get the scattering function S(E,T), or by considering Q-slices to study the S(Q,E,T).\\
On IN16B, the standard configuration with Si(111) backscattering crystals was applied. A neutron incident wavelength of 6.271 \AA was used, with an energy resolution of $\sim$ 0.7 $\mu$eV, and Q-range $\sim$ 0.2 - 2 \AA$^{-1}$. Elastic fixed window scans followed by inelastic scans, at 2 $\mu$eV, were performed.\\
The QENS measurements were extended to probe directly the time domain, in the ns range, on the neutron spin-echo spectrometer IN11. The IN11C 30$\degree$ detector option was used, with a neutron incident wavelength of 5.5 \AA. The instrumental resolution was measured at the base temperature
of about 2 K.
\subsection*{Vibrational Neutron Spectroscopy}
The neutron vibrational spectra presented in this work were obtained using both IN6 and IN1-Lagrange spectrometers (Institut Laue-Langevin, Grenoble).\\
On IN6, the INS spectra were collected up to 100 meV in the up-scattering, neutron energy-gain mode, at 250, 300 and 360 K, in the same way and under the same conditions as the QENS measurements, described above. The Q-averaged, one-phonon generalized phonon density of states (GDOS) \cite{gdosvdos1}, g$^{(n)}$(E), was obtained using the incoherent approximation \cite{incapp1, incapp2}. In the incoherent,
one-phonon approximation, g$^{(n)}$(E) is related to the measured scattering function S(Q,E) from INS by,

\begin{equation}
    g^{(n)}(E)=A\Big<\frac{e^{2W(Q)}}{Q^2}\frac{E}{n(E,T)+\frac{1}{2}\pm\frac{1}{2}}\Big>
\end{equation}

\begin{equation}
    g^{(n)}(E)=B\sum_i\Big(\frac{4\pi b_i^2}{m_i}\Big)x_ig_i(E)
\end{equation}
 
where A and B are normalisation constant. The + or - signs correspond to neutron energy loss or gain respectively. 2W(Q) is the Debye-Waller factor, and n(E,T) is the thermal occupation factor equal to $[exp(\frac{E}{k_BT}-1)]^{-1}$. The bra-kets indicate an average over the whole Q-range [28]. The neutron scattering length, mass, atomic fraction, and partial density of states of the i$^{th}$ atom in the unit cell are expressed in terms of b$_i$, m$_i$, x$_i$, and g$_i$(E), respectively. The weighting factors for various atoms in the units of barns.amu$^{-1}$ are \cite{booklet}: H: 81.37, D: 3.8, C: 0.46 and S: 0.032.\\
On the hot-neutron, inverted geometry spectrometer IN1-Lagrange, spectra were collected in the down-scattering, neutron energy-less mode at 10 K, with the fixed final analyzers energy of 4.5 meV. The incident energy was varied in a stepwise manner via Bragg scattering from copper or silicon monochromator crystals. In this work, using the doubly focused Cu(220) monochromator setting, the incident energy was $\sim$ 210 - 3500 cm$^{-1}$, leading after subtraction of the fixed final energy value (4.5 meV)  to an accessible energy transfer range of $\sim$ 180 - 3500 cm$^{-1}$. The low-energy part of the spectra was probed using the Si(111) and Si(311) monochromator settings, allowing to probe an energy transfer range up to 200 cm$^{-1}$.

\subsection*{Molecular Dynamics Simulations}
MD  simulations were performed  using  Gromacs-5.1.4 package,\cite{Berendsen1995,Lindahl2001,VanDerSpoel2005,Hess*2008,Pronk2013,Pall2015,Abraham2015} where a  leapfrog algorithm was adopted. Periodic  boundary  conditions  are  applied  in  all  directions. The Particle-mesh Ewald (PME) method is used for the electrostatic. Depending  on  the  ensemble  NVT  or  NPT,  we  used  a  velocity-rescaling  thermostat\cite{Bussi2007}(varying temperature, time constant 0.1 ps for equilibration and 0.5 ps for collection runs) and a Berendsen barostat (1 bar, compressibility 4.5.10$^{-5}$bar$^{-1}$, time constant 5 ps for equilibration and 2 ps for collection runs), respectively. We used the P3HT force field developed by Moreno \textit{et al.}\cite{Moreno2010a}. 100 chains of 20-mers RRa-P3HT were build with each dimer configuration chosen randomly according to the experimental regioregularity. The crystalline samples of RR-P3HT was build following the protocol explained in reference \cite{Poelking2014} in a 10x10 supercell. The energy was minimised using the steepest descent algorithm; through  energy  minimisation  using  the  steepest  descent  algorithm  the  convergence criterion was  set  such  that  the  maximum  force  is  smaller  than  10  kJ.mol$^{-1}$.nm$^{-1}$.
The temperature was stabilized by a NVT run for 5 ns and the pressure was stabilized by a NPT run for 50 ns. The amorphous samples were prepared from the melt as follow:
\begin{itemize}
    \item 100 chains were loosely and randomly packed using Packmol\cite{Martinez2009}
    \item the structures are relaxed  through  energy  minimisation  using  the  steepest  descent  algorithm.  The  convergence criterion was  set  such  that  the  maximum  force  is  smaller  than  10  kJ.mol$^{-1}$.nm$^{-1}$.
    \item NVT run at T=600K for 5 ns
    \item NPT run at T=600K and P=1bar for 50 ns
    \item NPT run at T=550K and P=1bar for 50 ns
    \item NPT run at T=500K and P=1bar for 50 ns
    \item ...
    \item NPT run at T=250K and P=1bar for 50 ns
    \item NPT run at T=200K and P=1bar for 50 ns
\end{itemize}
The size of the final boxes were checked to be larger than the length of an elongated  20mers  of  P3HT  (about 7 nm)  plus  the  cut-off  radius  used  for the van  der  Waals  forces (1.2 nm). For all the samples (crystalline or amorphous), a collection run was produced in NPT for 15 ns, and only the last 10 ns were used for the analysis.\\
The partial atomistic density of states were extracted from the MD by performing a FT of the velocity autocorrelation function (VACF), which were subsequently neutron-weighted \cite{gdosvdos1} to compare with the INS data.
\subsection*{Quantum Chemical Molecular and Periodic Calculations}
Both DFT-based isolated molecule and periodic quantum chemical calculations were performed, using Gaussian 16\cite{g16} and Castep \cite{Clark2005}, respectively, on different protonated and deuterated model structures of P3HT. \\
The model calculation in Gaussian consisted in applying the combined functional/basis-set b3lyp/6-311g(d,p) \cite{b3lyp} for the geometry optimization and the subsequent frequency and normal modes calculations.\\ 
Periodic calculations in Castep were performed using the plane-wave, norm-conserving \cite{ncpp} pseudopotentials, with a reciprocal space representation, within the generalized gradient approximation (GGA). The GGA was formulated by the Perdew-Burke-Ernzerhof (PBE) density functional \cite{pbe}. The Tkatchenko-Scheffler scheme \cite{TSvdw} was used to approximate a van der Waals dispersion correction. The break conditions for the self-consistent field (SCF) and ionic loops, for the relaxation of the atomic coordinates, were set to 10$^{-6}$ eV and 10$^{-2}$ eV/\r{A}, respectively. Phonon spectra were calculated within the framework of the linear response, density-functional perturbation theory \cite{DFPT}.\\
The simulated vibrational quantities, in terms the modes and their associated frequencies, were subsequently processed using aClimax \cite{aclimax} to facilitate a direct comparison with the measured INS spectra. 

\begin{acknowledgement}
The ILL is acknowledged for beamtime allocation on the spectrometers IN1-Lagrange, IN6, IN11 and IN16B, and on the diffractometer D16. A. A. Y. G. acknowledges EPSRC for the award of an EPSRC Postdoctoral Fellowship (EP/P00928X/1). J.N. acknowledges EPSRC for the grants EP/P005543/1, EP/K029843/1, and EP/J017361/1.
\end{acknowledgement}

\begin{suppinfo}
Synthesis procedure of h-RRa-P3HT and d-RRa-P3HT and associated NMR characterization. Quantitative NMR, XRay and UV-VIs absorption of all the samples. SANS data from D16, Q-dependent QENS data of IN6, fitting parameters of IN6 data, elastic and inelastic window scans from IN16B, EISF from IN11. Models of MD simulations. End-to-end distances, radius of gyrations, dihedral distributions as extracted from MD. Models for molecular and periodic DFT calculations.
\end{suppinfo}


\begin{mcitethebibliography}{63}
\bibitem[gdo()]{gdosvdos1}
The measured vibrational spectrum from INS is called a generalized density of
  states (GDOS). In contrast to the vibrational density of states
  \cite{gdosvdos2}, the GDOS involves a weighting of the scatterers (ions) with
  their scattering powers $\sigma$/M.\relax
\mciteBstWouldAddEndPunctfalse
\mciteSetBstMidEndSepPunct{\mcitedefaultmidpunct}
{}{\mcitedefaultseppunct}\relax
\EndOfBibitem
\end{mcitethebibliography}
%
\providecommand{\doi}
  {\begingroup\let\do\@makeother\dospecials
  \catcode`\{=1 \catcode`\}=2\doi@aux}

\end{document}